\documentclass[twocolumn]{aastex6}

\usepackage{amsmath, natbib}

\shorttitle{Near Infrared Variability of AGN in the COSMOS field}
\shortauthors{S\'anchez et al.}

\begin{document}

\title{Near Infrared Variability of obscured and unobscured X-ray selected AGN in the COSMOS field}

\author{P. S\'anchez\altaffilmark{1}; P. Lira\altaffilmark{1}, R. Cartier\altaffilmark{2}, V. P\'erez\altaffilmark{1}, N. Miranda\altaffilmark{3}, C. Yovaniniz\altaffilmark{1}, P. Ar\'evalo\altaffilmark{4}, B. Milvang-Jensen\altaffilmark{5}, J. Fynbo\altaffilmark{5}, J. Dunlop\altaffilmark{6}, P. Coppi\altaffilmark{7}, and  S. Marchesi\altaffilmark{8}}
\altaffiltext{1}{Departamento de Astronom\'ia, Universidad de Chile, Casilla 36D, Santiago, Chile.}
\altaffiltext{2}{Department of Physics and Astronomy, University of Southampton, Southampton, Hampshire, SO17 1BJ, UK}
\altaffiltext{3}{Departamento de Ciencias de la Computaci\'on, Universidad de Chile, Santiago, Chile.}
\altaffiltext{4}{Instituto de F\'isica y Astronom\'ia, Facultad de Ciencias, Universidad de Valpara\'iso, Gran Bretana No. 1111, Playa Ancha, Valpara\'iso, Chile}
\altaffiltext{5}{Dark Cosmology Centre, Niels Bohr Institute, University of Copenhagen, Juliane Maries Vej 30, 2100 Copenhagen, Denmark}
\altaffiltext{6}{Institute for Astronomy, University of Edinburgh, Royal Observatory, Edinburgh EH9 3HJ, UK}
\altaffiltext{7}{Yale Center for Astronomy and Astrophysics, 260 Whitney Avenue, New Haven, CT 06520, USA}
\altaffiltext{8}{Department of Physics \& Astronomy, Clemson University, Clemson, SC 29634, USA}

\begin{abstract}

We present our statistical study of near infrared (NIR) variability of X-ray selected Active Galactic Nuclei (AGN)  in the COSMOS field, using UltraVISTA data. This is the largest sample of AGN light curves in YJHKs bands, making possible to have a global description of the nature of AGN for a large range of redshifts, and for different levels of obscuration. To characterize the variability properties of the sources we computed the Structure Function. Our results show that there is an anti-correlation between the Structure Function $A$ parameter (variability amplitude) and the wavelength of emission, and a weak anti-correlation between $A$ and the bolometric luminosity. We find that Broad Line (BL) AGN have a considerably larger fraction of variable sources than Narrow Line (NL) AGN, and that they have different distributions of the $A$ parameter. We find evidence that suggests that most of the low luminosity variable NL sources correspond to BL AGN, where the host galaxy could be damping the variability signal. For high luminosity variable NL, we propose that they can be examples of ``True type II'' AGN or BL AGN with limited spectral coverage which results in missing the Broad Line emission. We also find that the fraction of variable sources classified as unobscured in the X-ray is smaller than the fraction of variable sources unobscured in the optical range. We present evidence that this is related to the differences in the origin of the obscuration in the optical and X-ray regimes.

\end{abstract}


\section{Introduction}\label{intro}

Active galactic nuclei (AGN) are one of the most energetic phenomena in the universe, and are characterized by their time-variable continuum flux in every waveband in which they have been studied. Variability studies are fundamental to understand the extreme physical conditions of accretion disks near supermassive black holes. The characteristic time scale of the variability ranges from days to years, and the variability magnitude is stronger in the X-ray, UV and optical ranges.

AGN are commonly classified in the optical range by the presence or absence of broad permitted emission lines ($\text{FWHM} \geq 2000$ km s$^{-1}$), into Broad Line AGN (or type I) and Narrow Line AGN (or type II), respectively. The unified model is one of the most successful approaches to explain this dichotomy. It postulates that a dusty torus around the central engine is the responsible for the different classes of AGN, which occur when we observe the source at different angles \citep{Antonucci85}. The most promising models include a clumpy torus and disk winds (see \citealt{Netzer15} and references therein), as they would explain the torus SED observed in the near infrared (NIR) and mid infrared (MIR) bands, and the existence of at least some ``Changing look'' AGN \citep{Tohline76,Shappee14,Denney14,LaMassa15,Ricci16}. Other objects might require more drastic modifications to the unified model, like those that seem to lack a Broad Line Region (BLR), called by some authors ``True type II'' AGN  \citep{Panessa02,Elitzur16} or Weak Emission Line Quasars (WLQ) \citep{Diamond-Stanic09,Shemmer09,Wu12,Shemmer15,Luo15}. Moreover, the relation between the optical and X-ray obscuration is still a debatable matter \citep{Merloni14,Burtscher16,Marchesi16b}. Variability analysis can improve our understanding of these issues, as observation at different wavelengths trace different structures.

There is significant evidence of a strong correlation between the X-ray, UV, and optical variability \citep{Uttley03,Arevalo08,Arevalo09,Breedt09,Breedt10,McHardy16,Troyer16,Buisson17}, where the emission at shortest wavelengths is driving the variations. Also, correlations between the near infrared (NIR) and optical bands has recently been established \citep{Suganuma06,Lira11,Lira15}. The rest frame optical emission comes primarily from the accretion disk, and the infrared emission comes mostly from the dusty torus \citep{Lira11,Lira15}. While the spectral energy distribution (SED) of an AGN  around the rest-frame wavelength of $\lambda \sim 1\mu m$ samples simultaneously two emission components, the accretion disk and the hottest part of the dusty torus. This has been confirmed both photometrically and spectroscopically \citep{Glass92,Landt11}. Besides, \citet{Lira15} found evidence of the presence of NIR emission from the accretion disk in the J and H bands observations of  MCG-6-30-15 by cross correlating optical and NIR light curves. 

AGN variability seems to be well described as a stochastic process \citep{Kelly09,Kelly14,Graham14,Simm16}. Power spectral density (PSD) analysis is a useful tool to analize the physical processes involved in the stochastic variability of AGN. The PSD measures the variability power per temporal frequency $\nu$. However, the irregular sampling of ground-based light curves complicates the PSD analysis. Previous analysis show that AGN power spectra are well described by a broken power law with a PSD $\propto 1/\nu^2$ after the break. \citep{Collier01,Czerny03,Kelly09}, which is consistent with Damped Random Walk (DRW) or autoregressive processes. \cite{Kelly14} showed that AGN light curves can also be well described by continuous--time autoregressive moving average (CARMA) models, which fully account for irregular sampling and measurement errors. From these models, the power spectra of AGN can have different shapes, depending on the CARMA parameters used to model the light curve. \cite{Simm16} showed that most of their sources (around 90 type I AGN) were best described by a CARMA(2,0) process (i.e., the PSD can be represented by a broken power law, which, after the break, has a slope different than -2), which means that most of their sources deviate from a simple DRW model.

To understand the physics behind AGN we need multi-wavelength variability studies, from which we can determine the contribution of the distinct components of the emission. But multi-wavelength projects are expensive and difficult to accomplish. Several authors have studied the optical, UV, and X-ray variability of AGN for a significant  number of sources (e.g.  \citealt{Cristiani96,Nandra97,Turner99,Giveon99,VandenBerk04, deVries05,Rengstorf06,Schmidt10,MacLeod11,PalanqueDelabrouille11,Lanzuisi14,Graham14,Cartier15,Simm16,Caplar17}). However, very little is known about the variability in the near infrared range. The study of AGN variability in the NIR is particularly difficult, since the contamination from the host galaxy in this wavelength region can be large \citep{Hernan-Caballero16}. Therefore, for most sources, the NIR variability of the central source is overshadowed by the emission from the galaxy. \citet{Neugebauer89} studied the NIR variability for individual quasars using a sample of 108 optically selected sources. They showed that only half of their sources have a high probability of been variable. \citet{Enya02a,Enya02b,Enya02c} analyzed the variability of 226 AGN, in the J, H and K' bands. Their work suggests that most AGN are variable in the near-infrared. However, both studies have the following limitations: small number of epochs, limited redshift coverage and only some specific classes of AGN were considered.

UltraVISTA \citep{McCracken12} is an ultra-deep, near infrared survey on the COSMOS field, using the 4-m VISTA survey telescope of the European Southern Observatory (ESO). UltraVISTA is repeatedly imaging the field in 5 bands (YJHKs and NB118), covering an area of 1.5 deg$^2$. The most recent UltraVISTA data release (DR3) corresponds to the first 5 years of observations. We used individual OB (Observation Block) stacks, which are images corresponding to 0.5 or 1 hour of exposure, to perform a near infrared AGN variability analysis. The advantages of these data compared to previous surveys is that UltraVISTA provides good quality and good resolution images (with a mean seeing of $\sim$ 0.8") at several epochs, and light curves with a length of almost 5 years and with good sampling. Besides, the depth of the images allowed us to cover a wide redshift range, therefore allowing the access to optical and near infrared rest-frame emission. 

In this work, we constructed light curves of known AGN selected from public catalogs of the COSMOS field \citep{Lusso12,Muzzin13,Marchesi16,Laigle16}, classifying them according to their X-ray, optical, and radio properties. We used these light curves to understand the differences in the variability behavior of the different classes of AGN, by implementing statistical tools widely used by the AGN community, like the Structure Function, continuous--time first-order autoregressive process or CAR(1), and the excess variance. 

This paper is organized as follows: in section \ref{data}, we briefly introduce the data set and the public catalogs used. In section \ref{class}, we give a definition of the different AGN classes considered in this work. In section \ref{lc} we describe the steps needed to get the light curves. In section \ref{analysis} we explain all the statistical tools used for the variability analysis. In section \ref{results} we present the results of the different analyses, considering the whole data set and the different classifications. In section \ref{discussion} we discuss the physical implications of our work and we summarize the main results. The photometry reported is in the AB system. We adopt the cosmological parameters $H_0=70$ km s$^{-1}$ Mpc$^{-1}$, $\Omega_m=0.3$ and $\Omega_\Lambda=0.7$.

\section{Data}\label{data}

\subsection{NIR data}\label{nirdata}

Our work is based on the near infrared (NIR) imaging data from the UltraVISTA survey \citep{McCracken12}, which has repeatedly imaged the COSMOS field during five years in the \textit{YJHKs} bands\footnote{The survey also uses the NB118 band \citep{Milvang-Jensen13}.}. The data considered in this work were taken between December 2009 and June 2014, using the VIRCAM instrument on the VISTA telescope at Paranal \citep{Emerson06, Dalton06, Sutherland15}. Further details of the data set can be found in \citet{McCracken12}.  

The final product of the UltraVISTA survey are stacked images and their corresponding weight maps for each filter band. However for our purpose, we worked with the individual OBs stacks, which have a total exposure of 0.5 or 1 hour, produced by the Cambridge Astronomy Survey Unit (CASU)\footnote{http://casu.ast.ac.uk/surveys-projects/vista/technical/data-processing}, and provided by the survey team, from which we have constructed our light curves. The reduction process done by CASU includes dark subtraction, flat-fielding, sky-subtraction, astrometric and photometric calibration, therefore, we did not  implement any reduction steps\footnote{see \citet{Emerson04,Irwin04,Lewis10,Milvang-Jensen13} for more details  of the CASU processing steps}, with the exception of photometric calibration (see section \ref{calibration}). The pixel scale of the images is 0.34"/px, and the average $5\sigma$ magnitude limits for the single images in each bands are 23.3, 23.1, 22.2, and 22.1 for the Y, J, H and Ks bands, respectively.

\subsection{Ancillary data}\label{catalogs}

We take advantage of the huge amount of ancillary data available for the COSMOS field, ranging from X-rays to radio waves. In our analysis we use four public catalogs. 

The first one is a public Ks-selected catalog of the COSMOS/UltraVISTA field using the UltraVISTA data \citep{Muzzin13}. This catalog provides photometry for 30 bands, covering the range 0.15-24 $\mu$m, besides other parameters related to the quality of the UltraVISTA photometry. For our work, the \textit{contamination} parameter resulted particularly important, as indicates whether an object's photometry has been contaminated by a nearby bright star. When the value of this parameter is zero there is no contamination.

The second one is the catalog of optical and infrared counterparts of the Chandra COSMOS-Legacy Survey \citep{Marchesi16}. This catalog contains 4016 X-ray sources from the 4.6 Ms Chandra program on 2.2 deg$^2$ of the COSMOS field \citep{Civano16}, with 3877 sources having an optical/IR counterpart. The catalog provides X-ray fluxes measured in three bands (Soft: 0.5-2 keV, Hard: 2-10 keV, and Full: 0.5-10 keV), hardness ratios, intrinsic neutral hydrogen ($N_H$) column densities, luminosity distances, identification and photometry of the counterparts in $i$, Ks and 3.6$\mu$m, spectroscopic redshift and classification for 1770 sources, photometric redshift and classification for 3885 sources, among other measurements. We complement the information of this catalog, with the new catalog of the 1855 extragalactic sources in the Chandra COSMOS-Legacy survey catalog having more than 30 net counts in the 0.5-7 keV band \citep{Marchesi16b}. This catalog provides new values of $N_H$ and the photon index ($\Gamma$) computed through spectral fitting.  We use the values of $N_H$ reported by  \cite{Marchesi16b} when their are available, otherwise we use the values reported by \cite{Marchesi16}.

Besides, we used the COSMOS2015 catalog \citep{Laigle16}, which provides photometric redshifts and stellar masses for more than half a million objects over 2 deg$^2$ in the COSMOS field, including photometry for several bands, with a wavelength coverage from the ultraviolet to the radio regimes. In particular we used SuprimeCam B band photometry \citep{Taniguchi07,Taniguchi15}, and VLA photometry at 1.4 GHz \citep{Schinnerer04,Schinnerer07,Schinnerer10,Bondi08,Smolcic14}.

Finally, \cite{Lusso12} provides measurements of Bolometric luminosities and Eddington ratios for X-ray selected broad-line (382 sources) and narrow-line (547 sources) AGN from the XMM--Newton survey in the COSMOS field \citep{Brusa10}. The bolometric luminosities are computed from the integrated SED.

The X-ray selection currently is the least biased but most expensive method to identify AGN, thus we considered as base for our analysis the \citet{Marchesi16} catalog, as this allow us to work with sources securely classified as AGN. We cross-matched this catalog with that of \citet{Muzzin13}, in order to secure the quality of the UltraVISTA photometry for every source, saving only the sources having a \textit{contamination} parameter equal to zero. Then, we cross-matched the catalog with the \citet{Laigle16} catalog, in order to obtain the B band and radio photometry. Finally, the resultant catalog was cross-matched with the \cite{Lusso12} catalog to obtain the bolometric luminosities for 718 of our sources. The final cross-matched catalog (hereafter the clean-AGN catalog) contains 3050 sources. 

\section{AGN Classification}\label{class}

Most of the variability studies available in the literature are based on objects classified as type I or unobscured AGN \citep{Cristiani96,Nandra97,Turner99,Enya02a,Enya02b,Enya02c,VandenBerk04,deVries05,Schmidt10,Lanzuisi14,Graham14,Simm16}. Since we have well sampled NIR light curves, we can expand our analysis and try to look for possible differences in variability between objects classified as obscured or unobscured according to different criteria. Using the information available from three public catalogs described in section \ref{catalogs}, we classified the sources of the clean-AGN catalog in the following way:

\subsection{Spectroscopic Classification}\label{spec_class}

\citet{Marchesi16} made use of the master spectroscopic catalog available for the COSMOS collaboration (M. Salvato et al., in preparation), to classify spectroscopically 1770 sources. For details on the source of the spectroscopic redshifts see \citet{Marchesi16}. They classified the objects according to the following criteria:

\begin{enumerate}

\item \textit{Broad line} (BL): sources with at least one broad ($\text{FWHM} \geq 2000$ km s$^{-1}$) emission line in their spectra (632 sources).
\item \textit{Not Broad line} (NL): sources that do not present broad lines (they show narrow emission lines or absorption lines). These sources have not been separated between star-forming galaxies or type II AGN because most of the sources have low S/N spectra, or are in an observed wavelength range which does not allow to use emission line diagnostic diagrams to separate Type II AGN and star-forming galaxies (1049 sources).
\item \textit{Stars}: sources spectroscopically identified as stars (89 sources).

\end{enumerate}

From these sub-samples, we have 563 BL and 952 NL in the clean-AGN catalog.

\subsection{Photometric Classification}\label{photo_class}

\citet{Marchesi16}  also provide photometric redshifts and classification for 3885 objects. They used the method described by \citet{Salvato11}, which  adjusts templates to the sources multiwavelength SEDs. The templates are divided in \citep{Salvato09}: `unobscured AGN', which corresponds to a type I AGN or type I QSO template (894 sources), `obscured AGN', which corresponds to a type II AGN or type II QSO template (365 sources), `galaxy', which corresponds to a elliptical, spiral, or starburst galaxy (2475 sources), and `star' (121 sources).  

From these sources, in the clean-AGN catalog we have 688 Unobscured AGN, 295 Obscured AGN, 1944 Galaxies, and 76 Stars.

\subsection{X-ray Classification}\label{xr_class}

Typically, AGN are classified according to their X-ray obscuration by using the Hardness Ratio: $HR = (H - S)/(H + S)$  (where H are the hard-band counts and S the soft- band counts, respectively), or by the intrinsic hydrogen ($N_H$) column density. Since our sample has a wide dynamic range in redshift, we decided to use the $N_H$ provided by \cite{Marchesi16b} (computed from spectral fitting) for the brightest sources, and by \cite{Marchesi16} (computed from the HR-z curve) for the faintest sources, instead of the $HR$, which is not corrected by redshift. We divided the sources between obscured and unobscured according to:

\begin{enumerate}

\item \textit{X-ray Unobscured AGN} (XR \textit{I} AGN): objects with $0 \leq N_H < 10^{22} $cm$^{-2}$
\item \textit{X-ray Obscured AGN} (XR \textit{II} AGN): objects with $N_H \geq 10^{22} $cm$^{-2}$

\end{enumerate}  

In the clean-AGN catalog we have 2114  type \textit{I} AGN and 936 type \textit{II} AGN.

\subsection{Radio Classification}\label{radio_class}

The COSMOS2015 catalog \citep{Laigle16} provides photometry in the SuprimeCam B band \citep{Taniguchi07,Taniguchi15}, and VLA fluxes at 1.4 GHz \citep{Schinnerer04,Schinnerer07,Schinnerer10,Bondi08,Smolcic14}. We used this information to classify our sources as radio-loud and radio-quiet using the ratio:

\begin{equation}\label{RL}
R=\frac{L_{\nu}(5 \, \text{GHz})}{L_{\nu}(4440 \, \text{\AA})}
\end{equation}

Where $L_{\nu}(5 \, \text{GHz})$ is the radio luminosity of the source measured at  $5 \, \text{GHz}$ and $L_{\nu}(4440 \, \text{\AA})$ is the B band luminosity. Since the emission in the radio and optical regimes comes from the same source, it does not matter if we use directly the flux instead of the luminosity. We needed to apply a K-correction to the photometry provided by the COSMOS2015 catalog, since the values needed to calculate $R$ have to be in the rest-frame. To do this, we considered that the radio and optical emissions follow a power-law like $F \propto \nu^{-0.8}$ and $F \propto \nu^{-0.44}$ respectively. Therefore, the final flux values used to determine $R$ are:

\begin{eqnarray}\label{Kcorr}
F_{\nu}(5 \, \text{GHz})_{ \text{rest}}=F_{\nu}(1.4 \, \text{GHz})_{ \text{obs}} \left( \frac{1.4}{5} \right)^{0.8} (1+z)^{-0.2}  & \nonumber \\
F_{\nu}(4440 \, \text{\AA})_{ \text{rest}}= F_{\nu}(4440 \, \text{\AA})_{ \text{obs}} (1+z)^{-0.56}  \quad &
\end{eqnarray}

In the COSMOS2015 catalog, not all sources have a detection at 1.4 GHz: for those sources without a detection reported, we assume that the measured flux corresponds to the $4\sigma$ upper limit  of 45 $\mu$Jy reported in \citet{Schinnerer10}.
We then use the following criteria to classify our sources:

\begin{itemize}

\item Radio-loud (RL): the source has a real detection at 1.4 GHz and $R \geq 10$
\item Radio-quiet (RQ): the source has a $R<10$, including upper limits and real detections at 1.4 GHz.

\end{itemize}

In the clean-AGN catalog we have  355 Radio-loud AGN and 566 Radio-quiet AGN.

\section{Light Curve Construction} \label{lc}

\subsection{Calibration and Photometry}\label{calibration}

For the construction of the light curves we used the time-resolved UltraVISTA images. These images were reduced using the CASU pipeline. The CASU stacks consist of 16 images, one per detector. Detector 16 is known for its unstable gain, therefore we did not consider it in our analysis. We further PSF homogenized the images and applied a photometric re-calibration to match the UltraVISTA DR3\footnote{\scriptsize{http://www.eso.org/sci/observing/phase3/data\_releases/uvista\_dr3.pdf}} catalogs. 

We perform a PSF homogenization, taking all the images of a certain filter band to a common seeing value, to avoid false variability detection, due to differences in seeing. Each image was convolved with a Gaussian of width equal to $\sqrt{(\sigma_0^2-\sigma^2)}$, where $\sigma_0$ is the width corresponding to some of the worst seeing conditions, and $\sigma$ is the width of each individual image. Since the behavior of the seeing in the four filters is similar, we selected a fixed value of $\sigma_0 \sim 1.0"$ (or 2.95 pixels) for all the data sets, and discarded all the images with a seeing worst than this value. The mean seeing value of the images is $\sim 0.82"$ for J, H, and Ks, and $\sim 0.89"$ for the Y band. We discarded the 17\%, 10\%, 8\%, and 9\% of the total images available for the Y, J, H, and Ks bands respectively. 

With these new images, we proceeded with the source detection and photometry, using the public package SExtractor \citep{sextractor}. We generate catalogs for every image in the four NIR bands. The catalogs contain for every source: position, aperture magnitude and flux using an aperture of 2", star classification (CLASS\_STAR, given by the shape of the source), and the FLAG parameter, which informs if an object is saturated or has been truncated at the edge of the image.  

It is well known that SExtractor underestimates the photometric errors when there are correleted pixels and when the objects are dimmer than the background. Since the PSF homogenisation produce correleted pixels, we calculated the photometric errors using a method similar to the one implemented by \cite{Gawiser06}. We estimated the photometric errors by placing 2000 random circular apertures of a certain size, and measuring the number of counts ($F$) inside the apertures, taking care to not overlap the apertures with sources detected by SExtractor. We then calculated the standard deviation of the number of counts in the apertures. We repeated this procedure changing the size of the apertures between 1 to 14 pixels in aperture. We then modeled the standard deviation of $F$ as $\sigma_N=\sigma_1a N^{b}$, where $N=n_{pix}^{1/2}$, $n_{pix}$ is the number of pixels in the aperture, $\sigma_1$ is the standard deviation for an aperture of 1 pixel, and $a$ and $b$ are the parameters of the model. For every single image, we calculated the values of $a$ and $b$, and then calculated the final photometric error for every source as $\sigma_{phot}=\sqrt{\sigma_{N}^2 + F/\text{gain}}$, with $\sigma_N$ the number of pixels used in the aperture of 2". The typical values of $\sigma_1$ are 1.0, 1.1, 3.5, and 2.0 for the Y, J, H and Ks bands, respectively. The values of $a$ and $b$ are $\sim 1.0$ and  $\sim 1.6$, respectively.  

Finally, we produced the calibrated catalogs. We used the catalogs generated by SExtractor, and cross-matched them with the DR3 catalog for the corresponding photometric band. We selected all the sources classified as stars by SExtractor (CLASS\_STAR $\geq 0.9$) and with good quality in their photometry (FLAG $=0$), and took the difference between their measured magnitudes and the magnitudes according to the DR3 catalogs for the 2" aperture. We then used a linear fit to model these residuals, following a procedure similar to \citet{Cartier15}, to prevent possible non-linearities in the detector, that might be produced after the PSF homogenization, with $m-m_{dr3} = \alpha + \beta\times m$, where $m$ is the magnitude in our catalogs for a certain band, and $m_{dr3}$ the magnitude provided by the DR3 catalogs. Finally, the calibrated magnitudes ($m_{cal}$) are computed as $m_{cal} = m - (\alpha + \beta\times m)$, and the calibrated errors as $\sigma_{cal}=\sqrt{\sigma_{phot}^2+\text{var}}$, where var is the variance of $(m-m_{dr3})-(\alpha + \beta\times m)$. The typical values of $\alpha$ are -0.55, -0.81, -1.42, and -1.91 for the Y, J, H and Ks bands, respectively. For $\beta$ we normally have a value of $\sim 0.0002$, $\sim 0.001$, $\sim 0.005$, and $\sim 0.009$ for the Y, J, H and Ks bands, respectively. Therefore, the non-linearities of the detectors are negligible.

\subsection{Light Curve Generation}\label{lc_gen}

We constructed light curves for all the sources in the clean-AGN catalog with a detection in the UltraVISTA single images. To construct the light curves, we cross-matched the clean-AGN catalog with every calibrated catalog, for which we knew their associated Julian dates, using a radius of 1". We discarded the outskirts of the images, considering only the regions with a distance greater or equal to 0.015 degrees to the border. We then constructed light curves for each source, saving only epochs where the SExtractor FLAG parameter was equal to zero. This prevents false detections of variability due to bad photometry. As a rule of thumb, we only saved those light curves with more than three epochs.

\begin{figure}
\begin{center}
\includegraphics[scale=0.55]{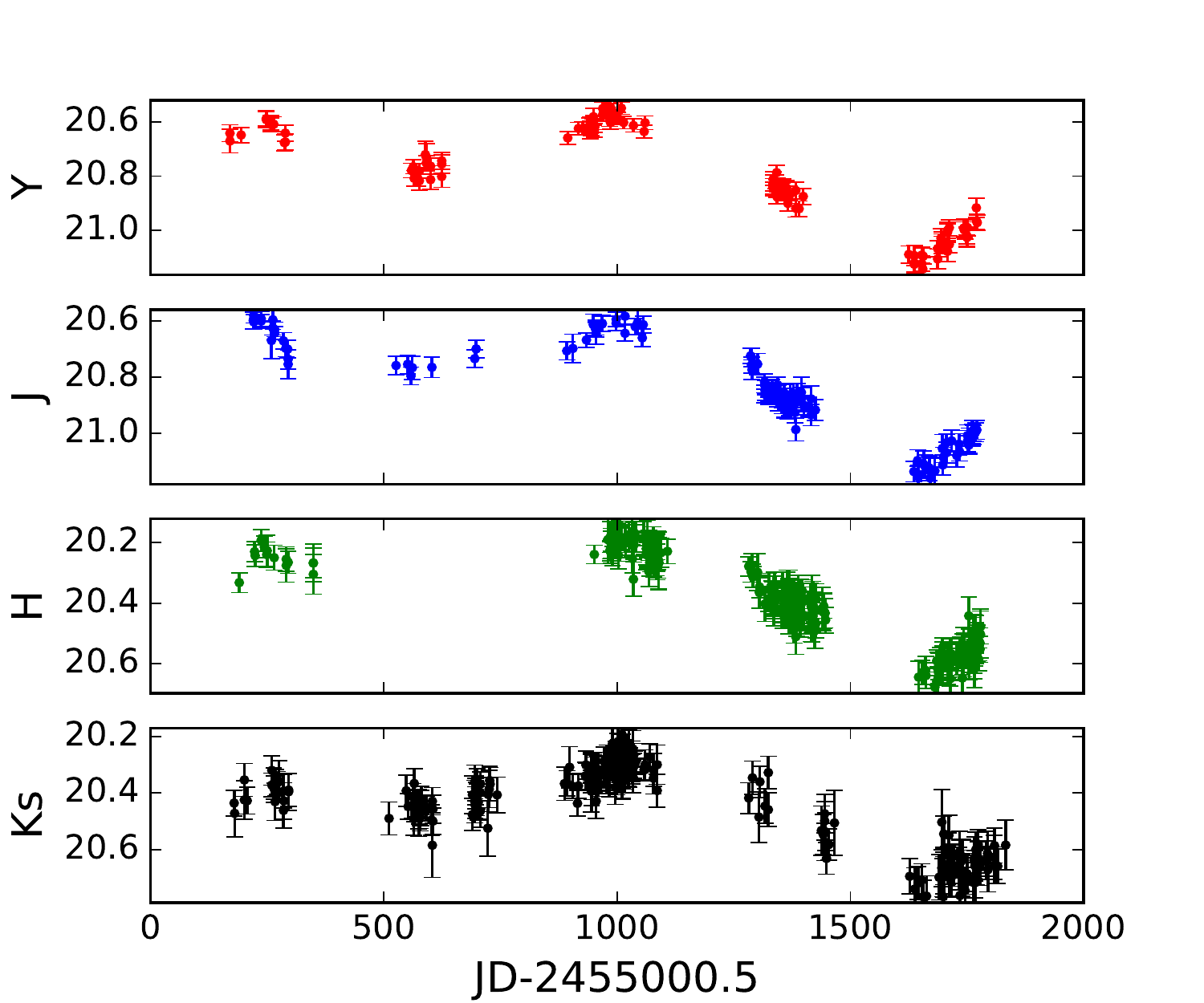}
\caption{Observed light curve for a BL - XR II - TypeI - RQ  AGN located at RA$=150.45187^{\circ}$ and DEC$=2.144811^{\circ}$, with ID cid$\_$543 from \cite{Marchesi16} catalog, located at z=1.298, variable in all the filter bands.\label{figure:lc_fig}}
\end{center}
\end{figure}

\begin{figure}
\begin{center}
\includegraphics[scale=0.55]{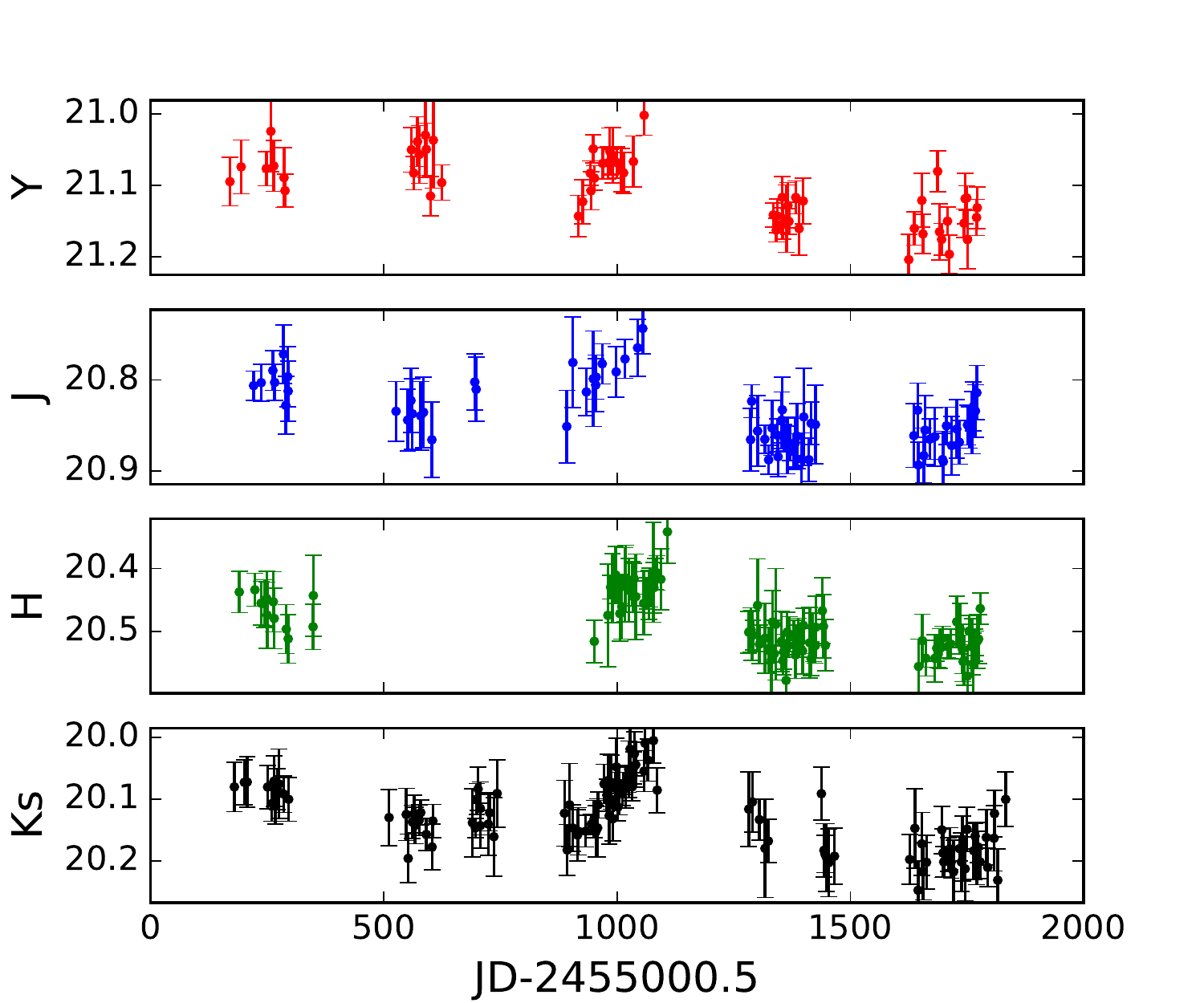}
\caption{Observed light curve for a NL - XR II - Galaxy - RQ AGN located at RA$=150.097790^{\circ}$ and DEC$=1.845247^{\circ}$, with ID cid$\_$254 from \cite{Marchesi16} catalog, located at z=0.711, variable in all the filter bands.\label{figure:lc_fig2}}
\end{center}
\end{figure}

\begin{figure}
\begin{center}
\includegraphics[scale=0.55]{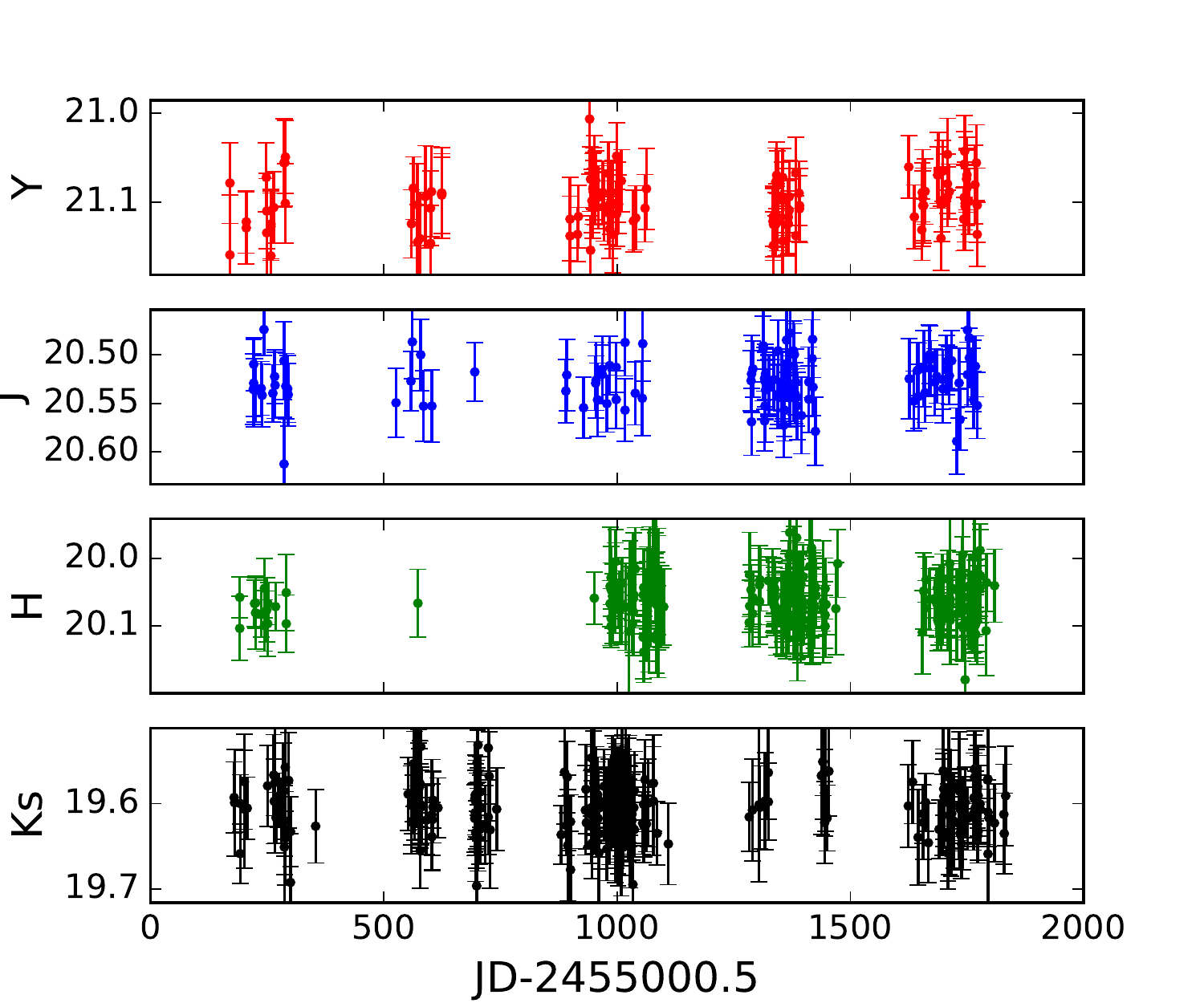}
\caption{Observed light curve for a  XR II - Galaxy - RL AGN located at RA$=149.43076^{\circ}$ and DEC$=1.939061^{\circ}$, with ID lid$\_$2414 from \cite{Marchesi16} catalog, located at z=0.916, non variable in all the filter bands.\label{figure:lc_fig3}}
\end{center}
\end{figure}

\begin{figure}
\begin{center}
\includegraphics[scale=0.43]{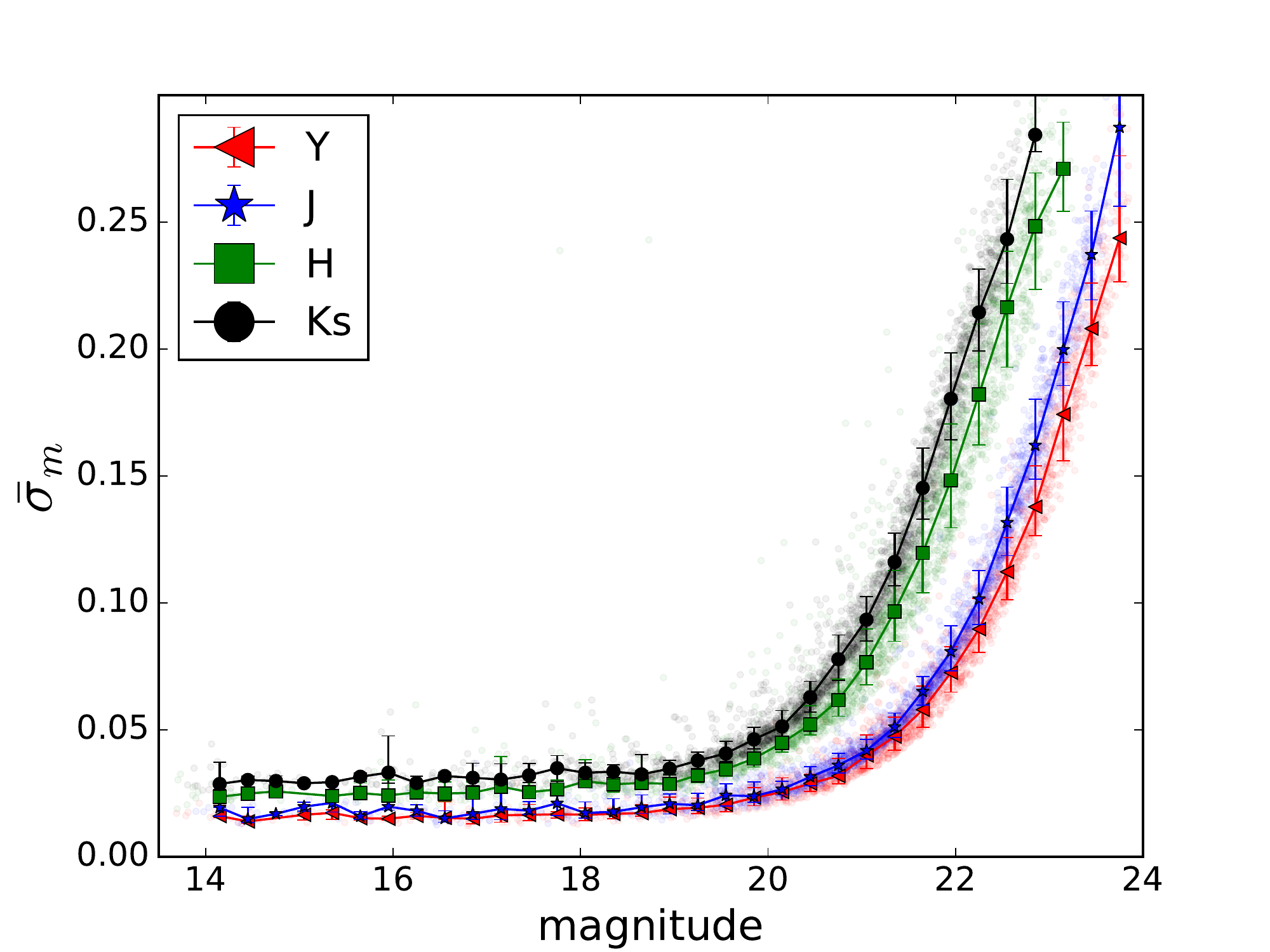}
\caption{Mean magnitude error vs mean magnitude for all the final light curves, for the four photometric bands Y (red), J (blue), H (green) and Ks (black).\label{figure:StoN}}
\end{center}
\end{figure}

After this preliminary construction, we cleaned every light curve following a $\sigma-$clipping procedure. First, all the epochs with magnitude error bigger than twice the mean magnitude error were rejected; second, we fitted an order five polynomial to the light curves, and rejected all the epochs with a distance from the polynomial bigger than $2\sigma$, with $\sigma^2=(\sigma_{epoch}^2+std)$, where $\sigma_{epoch}$ is the magnitude error in each epoch, and $std$ is the standard deviation of the whole light curve. Then, we only saved those light curves which ended with three epochs or more after the cleaning process. Finally, we transformed every light curve to the AGN rest-frame: $t_{rest}=t_{obs}/(1+z)$. Where $t_{rest}$ is the light curve time at the rest-frame in days and $t_{obs}$ is the observed time. Every variability feature (see section \ref{analysis}) was computed in the rest-frame of the AGN. We generated 1715, 1895, 1835, and 2107 light curves with magnitudes up to the 5$\sigma$ limit in the Y, J, H and Ks bands respectively. Figures \ref{figure:lc_fig} and \ref{figure:lc_fig2} show examples of variable light curves in the four bands for objects classified as a BL AGN and NL AGN. The light curves are plotted in the observed frame. Figure \ref{figure:lc_fig3} is an example of a non-variable light curve. Figure \ref{figure:StoN} shows the mean error in magnitudes vs the mean magnitude for the final light curves. From the figure we can see that the Y band has the best quality in the photometry, and the Ks band has the worst. This is expected, since as we move to redder bands, the sky brightness increases.

\begin{figure}
\begin{center}
\includegraphics[scale=0.4]{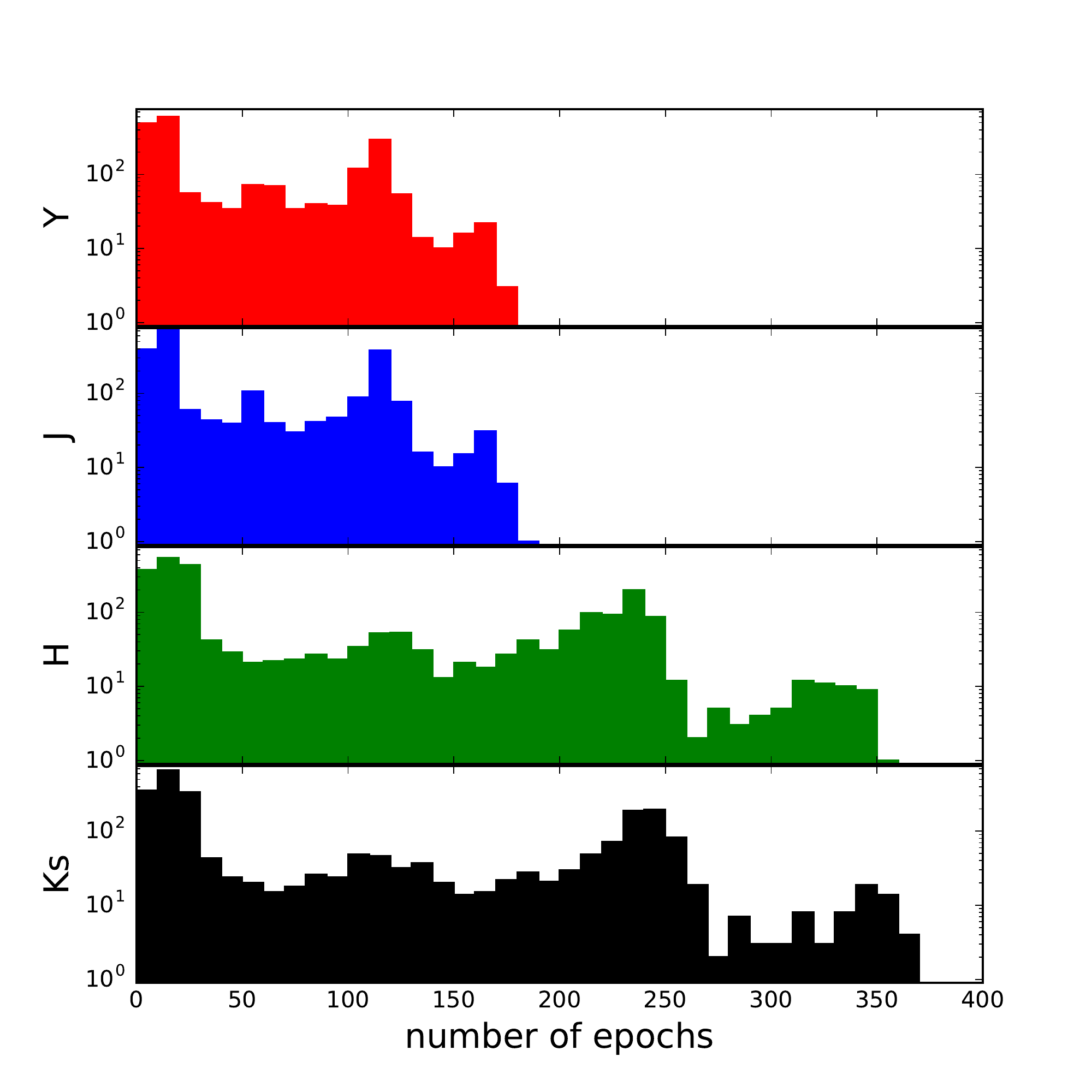}
\caption{Histogram in logarithmic scale of the number of epochs in the light curves for the four photometric bands Y (red), J (blue), H (green) and Ks (black).\label{figure:hist_epochs}}
\end{center}
\end{figure}

\begin{figure}
\begin{center}
\includegraphics[scale=0.4]{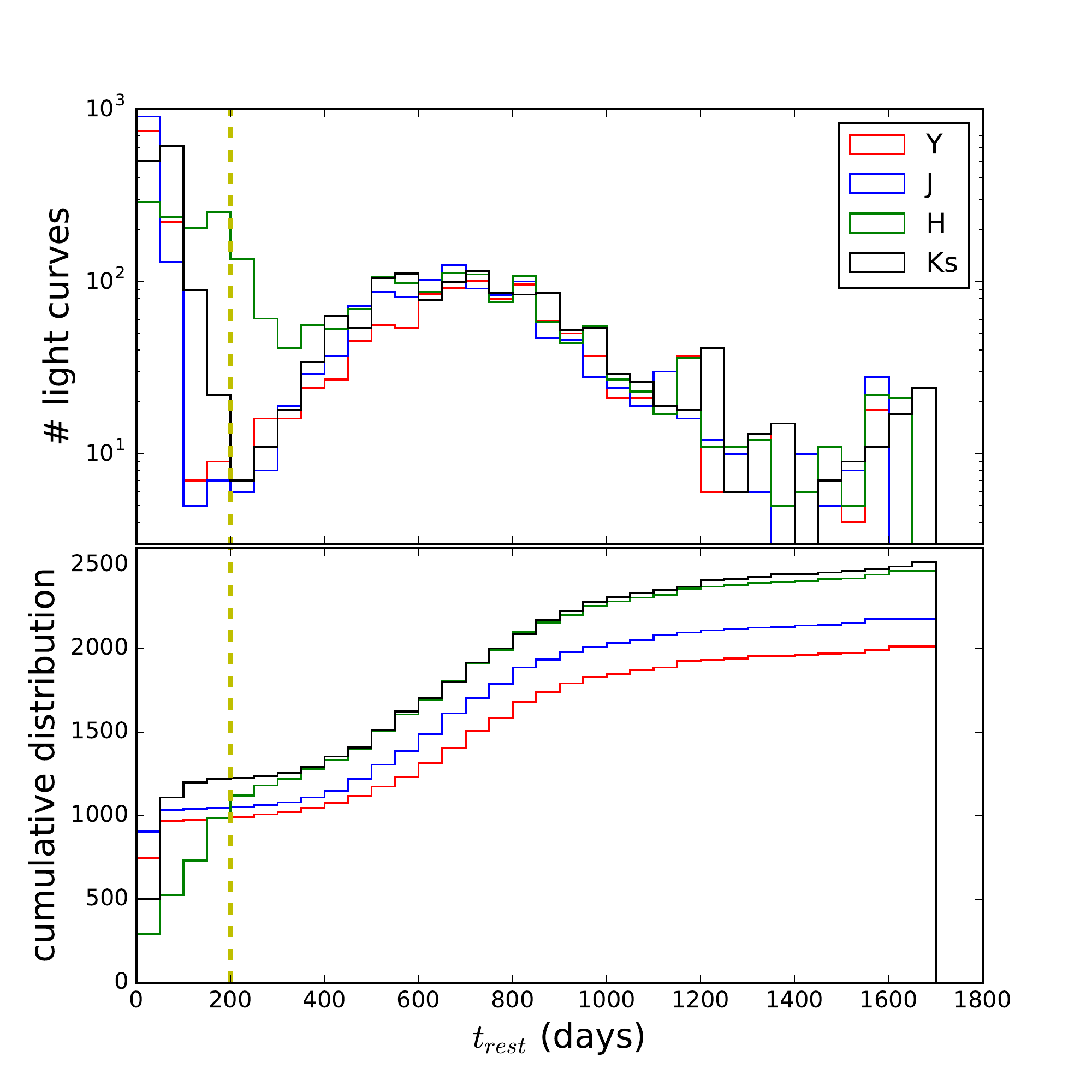}
\caption{Top: Histogram in logarithmic scale of the rest frame time length  ($t_{rest}$), defined as the light curve's observed length divided by $(1+z)$, for the four photometric bands Y (red), J (blue), H (green) and Ks (black). Bottom: Cumulative distribution of the rest frame time length. The yellow dashed vertical line marks $t_{rest}=200$.\label{figure:hist_timerange}}
\end{center}
\end{figure}

The UltraVISTA survey is split into two sets of strips called deep and ultra-deep (see \citealt{McCracken12}). In the first year of UltraVISTA both sets of strips were observed, thereby providing a nearly homogeneous coverage of the field. In the following four years only the ultra-deep strips were observed. This clearly influences both the number of epochs and the total length of the light curves for the objects in this study. Additionally, each strip consists of 3 pointings (pawprint positions),  where each OB obtained images were jittered around one such pawprint position. Therefore,  we have light curves with a number of epochs that ranges from 3 to 365. Furthermore, the UltraVISTA project did not observe the COSMOS field in every photometric band by the same number of epochs. The band with the best sampled light curves is the Ks band, with a mean number of 92 epochs, followed by the H band, with 85 epochs, the J band with 48 epochs and finally the Y band with 45 epochs, in average. Figure \ref{figure:hist_epochs} shows the number of epochs in the light curves. Most of the light curves have less than 50 epochs. Figure \ref{figure:hist_timerange} shows the rest frame time length  ($t_{rest}$), defined as the length of the light curve observed length divided by $(1+z)$. From the figure we can see that half of the light curves have a length of less than 200 days.

\section{Variability Analysis} \label{analysis}

To characterize the variability of the sources we used $P_{var}$, the excess variance ($\sigma_{rms}$), the Structure Function and the continuous--time autoregressive process or CAR(1). The first 2 methods are used to detect the  variability of a source, without taking into account the shape of the variation, whilst the last 2 methods can characterize the structure of the variability. 

\subsection{The $P_{var}$ parameter}\label{Pvar}

The $V$ parameter is defined by \citet{McLaughlin96} as a value related to the probability of a source to be variable, and is used in different variability studies (e.g. \citealt{Paolillo04,Young12,Lanzuisi14,Cartier15}). It is defined from the $\chi^2$ of the light curve:
\begin{equation}\label{chi2}
\chi^2=\sum_{i=1}^{N_{obs}}\frac{(x_i-\bar{x})^2}{\sigma^2_{err,i}}
\end{equation}
where  $x_i$ is the magnitude at each epoch, $\sigma_{err,i}$ is its error, $\bar{x}$ is the mean magnitude, and $N_{obs}$ is the number of epochs in which the object was detected. If the source were intrinsically non-variable, the value of $\chi^2$ would be $\sim (N_{obs}-1)$, the number of degrees of freedom in the data. To see whether the $\chi^2$ value is consistent with real variability, we calculate the probability $P_{var}=P(\chi^2)$ that a $\chi^2$ lower or equal to the observed value could occur by chance for an intrinsically non-variable source. If the value of $P_{var}$ is large, $1-P_{var}$, the probability that the variability observed is due to Poisson noise alone and the source is intrinsically non-variable is low. Therefore we say that $P_{var}$ corresponds to the probability that the source is intrinsically variable. We then define the variability index as $V = -\text{log}(1-P_{var})$. Following \citet{Lanzuisi14} and \citet{Cartier15}, we defined $P_{var} \geq 0.95$ (or $V\geq1.3$) as our threshold to define a variable object. Therefore, we might expect to detect a 5\% of false positive sources.

\subsection{The Excess Variance $\sigma^2_{rms}$}\label{ExVar}

The normalized excess variance $\sigma^2_{rms}$  \citep{Nandra97,Turner99,Allevato13,Lanzuisi14,Cartier15}, is a measure of the variability amplitude. Here we adopt the definition: 

\begin{equation}\label{exvar}
\sigma^2_{rms}=\frac{1}{N_{obs}\bar{x}^2}\sum^{N_{obs}}_{i=1}[(x_i-\bar{x})^2-\sigma^2_{err,i}]
\end{equation}

And its error due to Poisson noise is:

\begin{equation}\label{errexvar}
err(\sigma^2_{rms})=\frac{S_D}{\bar{x}^2 N_{obs}^{1/2}},
\end{equation}

\begin{equation}\label{S_D}
S^2_D=\frac{1}{N_{obs}}\sum^{N_{obs}}_{i=1}\{[(x_i-\bar{x})^2-\sigma^2_{err,i}]-\sigma^2_{rms}\bar{x}^2\}^2
\end{equation}

For a non-variable object, the excess variance can be negative: if variability is not detected, due to large errors, the value of $err(\sigma^2_{rms})$ can be greater than $\sigma^2_{rms}$. The excess variance has to be treated with extreme care \citep{Allevato13}, since it can be biased by the structure of the variability of the source, the sampling and the length of the light curve. However, we only used the parameter to say if a source has detected variability when $(\sigma^2_{rms}-err(\sigma^2_{rms}))>0$, i.e., when the intrinsic amplitude of the variability (corrected by the photometric errors) is greater than zero. If a source accomplishes this criteria and has $P_{var} \geq 0.95$ we say that the object is intrinsically variable. This approach helps to reduce the fraction of false positives detected when we only consider the $P_{var}$ parameter to classify an object as variable. 

\subsection{The Structure Function}\label{SF}

The Structure Function describes the variability of a source by quantifying the amplitude of the variability as a function of the time lapse between compared observations ($\tau$) \citep{Cristiani96,Giveon99,VandenBerk04, deVries05,Rengstorf06,Schmidt10,PalanqueDelabrouille11,Graham14,Cartier15,Kozlowski16a,Caplar17}. \cite{Kozlowski16a} provides a good summary of the different definitions of the Structure Function used in the literature. In particular he proposed to define it as: $\text{SF}_{obs}(\Delta t)=0.741\times \text{IQR}$, where IQR is the interquartile range between 25$\%$ and 75$\%$ of the sorted $y(t) - y(t + \Delta t)$ distribution. He also provides a method to measure the Structure Function taking into account the photometric noise: $\text{SF}^2_{true}(\Delta t)=0.549( \text{IQR}^2(\Delta t) - \text{IQR}^2(n))$, where $\text{IQR}(n)$ is the interquartile range between 25$\%$ and 75$\%$ of the sorted $\Delta m$ for $\Delta t < 2$ days.

The Structure Function of AGN is generally well described by a broken power law. The time scale were the break happens is known as the decorrelation time scale. \cite{Emmanoulopoulos10} discussed the different limitations of the use of the decorrelation time scale as a physically meaning parameter, which can be severely affected by the light curve sampling. Therefore, we limit our analysis to the regime were we can describe the Structure Function as a single power law (i.e. before the break):

\begin{equation}\label{SF_eq}
\text{SF}(\tau)=A\left( \frac{\tau}{1\text{yr}}\right)^{\gamma} 
\end{equation}
where $A$ corresponds to the mean magnitude difference on a one year time-scale and $\gamma$ is the logarithmic gradient of this change in magnitude. 

The $\gamma$ parameter is directly related to the power spectral density (PSD) slope. When $\text{SF} \propto t^{\gamma}$ then $\text{PSD} \propto 1/ f^{1+2\gamma}$, therefore, the value of $\gamma$ changes depending of the type of process involved in the variation. For example, for a white noise process $\gamma=0$, and for a random walk process $\gamma=0.5$.

\citet{Schmidt10} provides two methods to define the Structure Function, the first one is similar to other definitions used in the literature: 

\begin{equation}\label{sfSch10}
\text{SF}(\tau)=\frac{1}{N_{bin}}\sum_{i,j} \left(  \sqrt{\frac{\pi}{2}} |\Delta m_{ij}| - \sqrt{\sigma_i^2 + \sigma_j^2}  \right) 
\end{equation}

Where the average $|\Delta m_{ij}|$ is taken over all the epoch pairs $i,j$ whose lag in time $\Delta t_{ij}=t_i-t_j$ falls inside the bin [$\tau - \frac{\Delta \tau}{2}$, $\tau + \frac{\Delta \tau}{2}$]. $N_{bin}$ corresponds to the number of pairs inside the bin, $\Delta m_{ij}$ is the difference in magnitude between the two epochs ($m_i-m_j$), and $\sigma_i$ and $\sigma_j$ are the magnitude errors for each epoch, respectively.  

The second method proposed by \citet{Schmidt10} is a Bayesian approach, where they model the Structure Function with a power-law using a Markov Chain Monte Carlo (MCMC) method. In this method, a list with all the possible epoch pairs is constructed. Then the Structure Function is modeled considering a likelihood $\mathcal{L}(A,\gamma)$, and priors $p(A)$ and $p(\gamma)$ defined as:

\begin{equation}\label{likelihood}
\mathcal{L}(A,\gamma)=\prod_{ij}\frac{1}{\sqrt{2\pi \text{V}_{\text{eff},ij}^2}} \text{exp}\left( -\frac{\Delta m_{ij}^2}{2\text{V}_{\text{eff},ij}^2}  \right) 
\end{equation}

\begin{equation}\label{Veff}
\text{V}_{\text{eff},ij}^2=\left[ A \left( \frac{\Delta t_{ij}}{1 \text{yr}}\right) ^{\gamma} \right]^2 +(\sigma_i^2 +\sigma_j^2) 
\end{equation}

\begin{equation}\label{priors}
p(A)\propto \frac{1}{A},  \quad  p(\gamma)\propto \frac{1}{1+\gamma^2}
\end{equation}

The main advantage in determining the Structure Function with a Bayesian approach compared with traditional definitions, is that it can avoid problems given by the sampling of the light curve and the selection of the bin size and shape, since these parameters are inferred directly from the data. This method is also less susceptible to windowing effects, given by the finite length of the light curve. Moreover, from the posterior distribution of the parameters, we can determine the mean value and the $1 \sigma$ errors of the measurements. In our particular case, the values of $\gamma$ and $A$ were constrained to be in the ranges $\gamma\in [0,10]$ and $A \in [0,1]$ mag/year. Besides, we only considered epoch pairs with a maximum separation of 1 year in the rest frame, in order to avoid the regime of the Structure Function after the break.

\begin{figure*}
\gridline{\fig{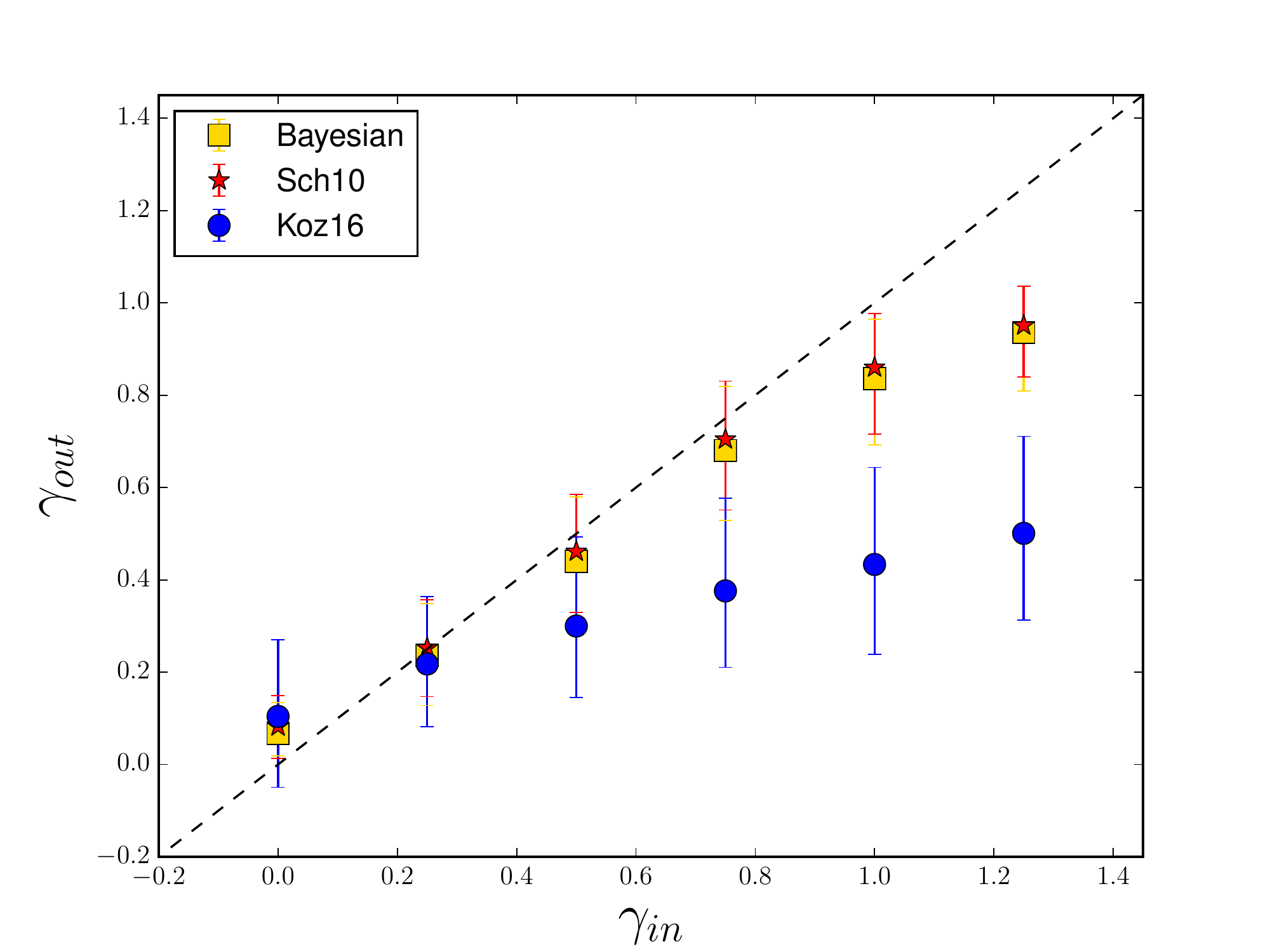}{0.5\textwidth}{}
          \fig{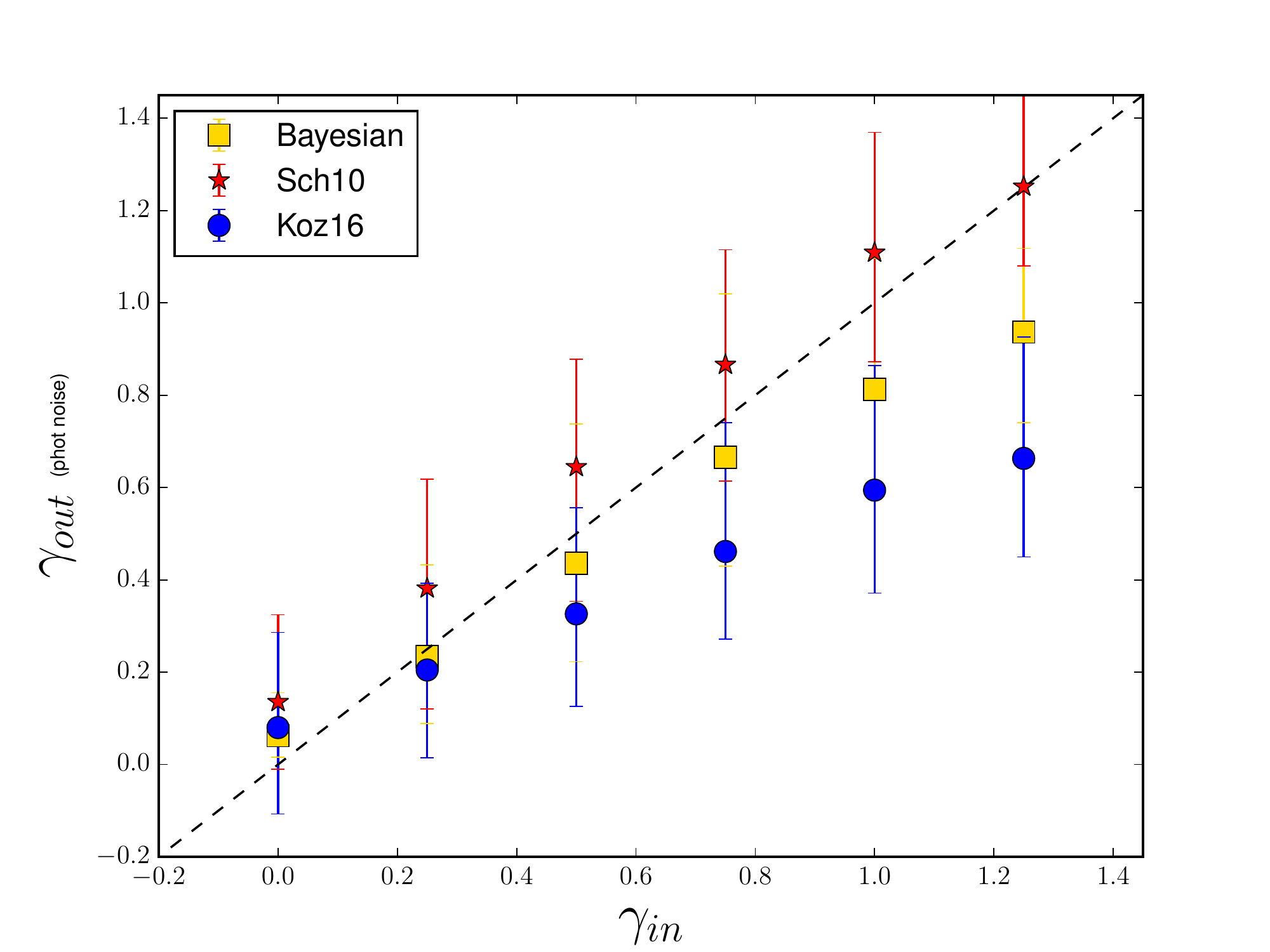}{0.5\textwidth}{}
          }
\caption{Results of the Structure Function analysis for the artificial light curves with power-law PSD. In the x-axis we show the inpunt value of $\gamma$ and in the y-axis we show the value of $\gamma$ computed by the diferent methods. The error bars in the y-axis correspond to the 15.86 and 84.14 percentiles of the output $\gamma$ distributions. The blue circles correspond to the results obtained using the definition given in \cite{Kozlowski16a}, the red stars correspond to the first definition given by \citet{Schmidt10} (eq. \ref{sfSch10}), and the yellow squares correspond to the Bayesian method proposed by \citet{Schmidt10}. The left panel corresponds to light curves without photometric noise. The right panel correspond to light curves with a photometric noise of 0.02 magnitudes. Light curves with ``NAN'' values of the    $\gamma$ parameter are not included. \label{figure:SF_test}}
\end{figure*}

In order to test which of the previously mentioned methods is more suitable for the analysis of our data, we simulated artificial light curves with a power-law PSD, and with a sampling representative of the UltraVISTA light curves in the Y band. To simulate the light curves we used the algorithm proposed by \cite{Timmer95}. Then, we analyzed which method recovers the slope of the PSD more accurately. We simulated light curves with $\gamma=[0, 0.25, 0.5, 0.75, 1, 1.25]$. We generated 1000 light curves without photometric noise and 1000 light curves with a photometric noise of 0.02 magnitudes (representative of our light curves in the Y band) for every value of $\gamma$. Figure \ref{figure:SF_test} shows the results of this analysis. The left panel of the figure shows the results for the light curves without photometric noise. For all these light curves we were able to obtain measurements of the $\gamma$ parameter by using the three methods. However it can be seen that the best results are obtained with the two methods proposed by \citet{Schmidt10}. The right panel of Figure \ref{figure:SF_test} shows the results for the light curves with photometric noise. We only were able to obtain measurements of the Structure Function parameters for all light curves with the method of \cite{Kozlowski16a} and with the Bayesian method of  \citet{Schmidt10}. For the case of the analytic definition of the Structure Function proposed by \citet{Schmidt10} (Eq.\ref{sfSch10}),  $\sim 70 \%$ of the light curves return ``NAN'' values of $\gamma$. These light curves are not considered in the results shown with red stars in the right panel of Figure \ref{figure:SF_test}. We think this is a consequence of a too high subtraction of the noise term in Eq. \ref{sfSch10}, as previously pointed out by \cite{Kozlowski16a}. Therefore, since the Bayesian method proposed by \citet{Schmidt10} gives the best results and is more stable under the presence of photometric noise, we decided to use this method in the rest of our analysis. All the results presented below for the Structure Function were computed using the Bayesian method.

\subsection{Continuous Time Autoregressive Process}\label{car}

 The light curves of AGN can be described by stochastic processes. In particular \citet{Kelly09} proposed that a continuous--time first-order autoregressive process or CAR(1) can be a good descriptor of this kind of variation. This model is also called ``Damped Random Walk", since it is represented by a stochastic differential equation which includes a damping term that pushes the signal back to its mean: 
\begin{equation}\label{car_eq}
dX(t)=-\frac{1}{\tau}X(t)dt+\sigma\sqrt{dt}\,\epsilon(t)+b\,dt,    \quad      \tau,\sigma,t>0 
\end{equation}
here, $X(t)$ is the AGN light curve, represented by the observed magnitude, $\tau$ is the ``relaxation time" of the process or the characteristic time for the time series to become roughly uncorrelated (related with the decorrelation time mentioned in the previous section), $\epsilon(t)$ is a white noise process with zero mean and variance equal to 1, and $\sigma$ is the amplitude of the variability on short time-scales compared with $\tau$. The mean value of the process is $b\tau$ and its variance is $\tau\sigma^2/2$.
From the fit of a DRW process, we can derive the power spectral density of the light curve, avoiding the windowing effect that appears when the PSD is derived directly from the data:

\begin{equation}\label{psd_eq}
\text{PSD}(f)=\frac{2 \sigma^2 \tau^2}{1+(2\pi \tau f)^2}
\end{equation}
where f is the frequency measured in days$^{-1}$.

There are two main regimes for the PSD$(f)$ of a DRW process. For short time-scales (compared to the relaxation time, i.e., $f\lesssim (2\pi \tau)^{-1}$), the power spectrum falls of as $1/f^2$. On longer time-scales than the relaxation time, the power spectrum flattens to white noise. Therefore, $\tau$ can be considered as the characteristic time--scale of the variability. \citep{Kelly09}.

Several authors claimed that the AGN light curves can be modeled in a better way by continuous--time autoregressive moving average (CARMA) models \citep{Kelly14,Simm16}. These models fully account for irregular sampling and measurement errors. CARMA models are generated by adding higher order derivatives to the stochastic differential equation given in Eq. \ref{car_eq}. \citet{Kelly14} provides a public  {\scriptsize \textsc{PYTHON}}  package to fit CARMA models called \textit{carma\_pack}\footnote{https://github.com/brandonckelly/carma\_pack}. The package includes the option to model DRW processes.

\begin{figure}
\begin{center}
\includegraphics[scale=0.6]{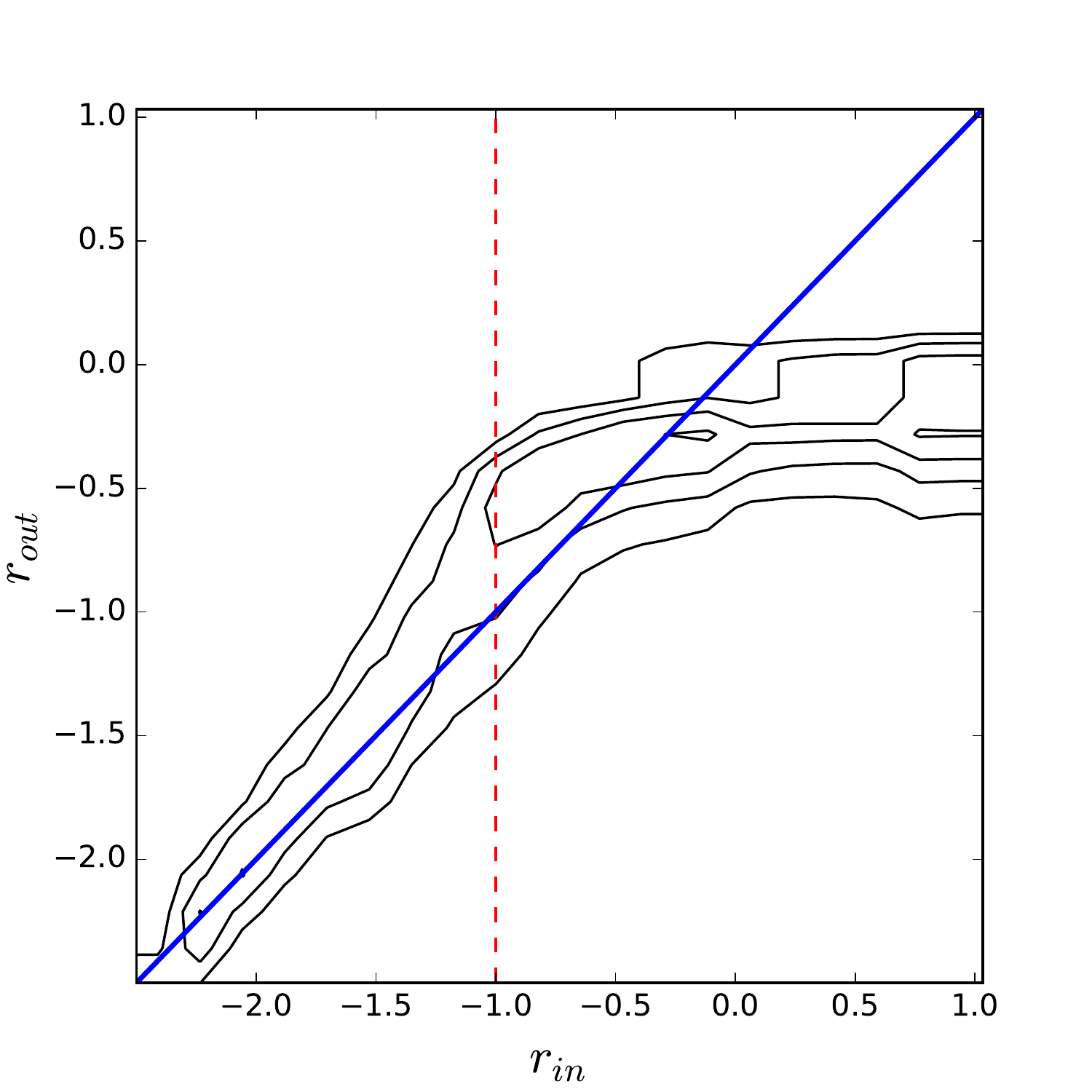}
\caption{Comparison of $r_{in}$ vs $r_{out}$ for simulated DRW light curves. The blue line shows the 1:1 relation. The red dotted line shows the region where the length of the light curve is 10 times the input $\tau$.\label{figure:tau_test}}
\end{center}
\end{figure}

\cite{Kozlowski17} presents an analysis of the limitations of the DRW to model AGN light curves. He demonstrated that it is necessary to have light curves with at least 10 times the length of the``relaxation time" in order to have accurate variability parameters derived from the DRW analysis. We tested whether this effect is present when we use \textit{carma\_pack} to fit DRW models to our light curves. We simulated DRW light curves with a sampling representative of the light curves in the Y band, following the approach proposed by \cite{Kelly09} using different values of $\tau$, and then we fitted these light curves using \textit{carma\_pack}. We compared the output $\tau$ obtained from the method with the input $\tau$ used to generate the light curves. We define the parameter $r=\text{log}_{10}(\tau/t_{lc})$, where $t_{lc}$ is the length of the light curve. Figure \ref{figure:tau_test} shows a comparison of $r_{in}$ calculated using the input $\tau$ and $r_{out}$ calculated using the output $\tau$. In the figure we confirm the results of \cite{Kozlowski17}. For light curves whose input $\tau$ is less than 10 times the length of the light curve, the output ``relaxation time'' is close to the original value. For the remaining light curves, the length is too short to give confident results. Since our light curves have a typical length of 5 years, we would be able to detect accurately values of $\tau$ lower than $\sim 180$ days. After correcting by redshift, the length of most of our light curves is too short to measure $\tau$ accurately. Therefore we decided not to include the DRW analysis in our results.

\section{Results}\label{results}

We define a source as intrinsically variable, when its light curve has $P_{var}>=0.95$ and $(\sigma^2_{rms}-err(\sigma^2_{rms}))>0$  (see section \ref{Pvar} and \ref{ExVar}). We found that $13.47\%$, $11.13\%$, $5.4\%$  $6.22\%$ of the total number of sources in the Y, J, H and Ks bands, respectively, are variable. However, as seen in Figures \ref{figure:hist_epochs} and \ref{figure:hist_timerange}, there are several light curves with poor sampling. Therefore, in order to homogenize our analysis, we only considered well sampled light curves, that is, light curves with at least 20 epochs and with a rest frame time length ($t_{rest}$) greater or equal to 200 days. Besides, we only considered sources with either spectroscopic or photometric redshifts available (we use the best redshift reported by \citealt{Marchesi16}), and with a total Ks magnitude (equivalent to SExtractor `AUTO\_MAG') brighter than 22.0. We call this downsized sample the ``clean-sample".

In this section we present the results of our variability analysis. Table \ref{tab:summary_lc} summarizes the number of light curves available for each object type in each NIR band before and after downsizing. The first value of every entrance in Table \ref{tab:summary_lc} corresponds to the results for the clean-sample. The second value gives the numbers for the whole data set, before we downsized the sample. We found that $27.4\%$, $21.7\%$, $9.1\%$ and $11.5\%$ of the sources from the clean-sample in the Y, J, H and Ks bands, respectively, are variable. As can be seen from the Table \ref{tab:summary_lc}, we only miss a small fraction ($\sim 10 \%$) of variable sources after the sample is downsized, while the relative fraction of variable sources doubles. 

The Y and J bands have the best quality in the photometry (see Figure \ref{figure:StoN}), and the larger fraction of variable sources. We will focus on the analysis of the clean sample in these bands in the following sections, unless otherwise noticed. 

\begin{figure}
\begin{center}
\includegraphics[scale=0.54]{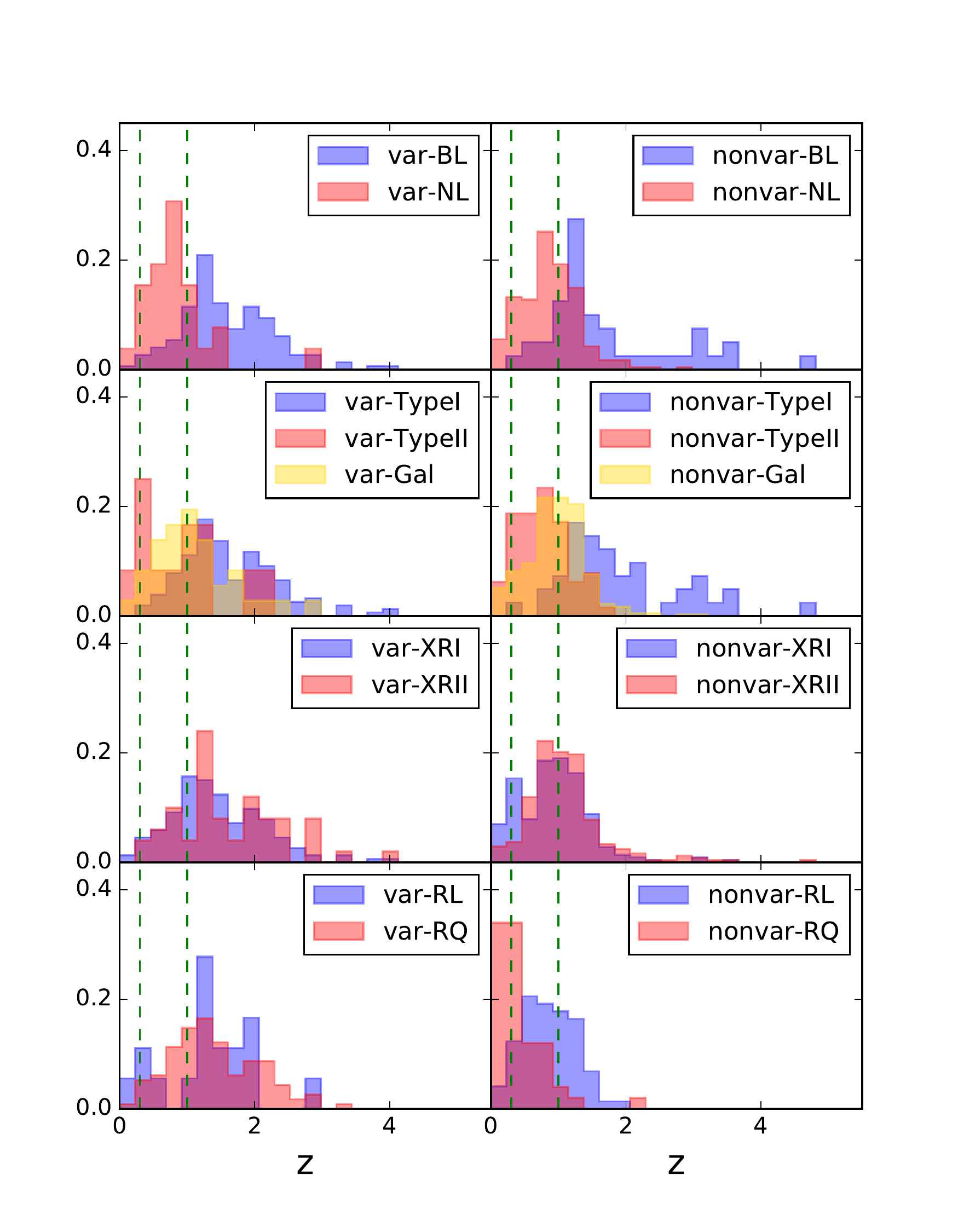}
\caption{Normalized histogram of the redshift distribution of the variable and non variable sources in the Y band for the four classes defined in Table \ref{tab:summary_lc}. The green dashed lines show the 0.3 to 1 redshift bin. From top to bottom: spectroscopic, photometric, X-ray and radio classifications.\label{figure:hist_redshift}}
\end{center}
\end{figure}

Figure \ref{figure:hist_redshift} shows the redshift distribution of the variable and non-variable sources for the different classifications in the Y band. From the figure we can see that there are not important differences in the distributions of the variable and non variable sources, except for the radio classification, where we can see that the non-variable sources are clustered at lower redshifts. This is produced by the spectroscopic classification of the sources with radio classification, since most of the non-variable sources are NL and are located at low redshifts, and most of the variable sources are BL and are located at higher redshifts.

\begin{table*}
\renewcommand{\thetable}{\arabic{table}}
\centering
\caption{Number of well sampled light cuves before and after downsizing: clean-sample / whole data set. In brackets we show the number of variable sources.}\label{tab:summary_lc}
\begin{tabular}{cccccc}
\tablewidth{0pt}
\hline
\hline
Classification & { } & Y & J & H & Ks\\
\hline
\hline
Spectroscopic &BL    &    196 / 432  & 203 / 447 & 207 / 425 & 223 / 470    \\ 
{ }&{ } & (152 / 162) & (129 / 141) & (64 / 67) & (79 / 83)  \\
{ }&NL  &      319 / 677 & 343 / 703  & 405 / 717 & 388 / 781   \\ 
{ }&{ } & (28 / 31) & (29 / 33) & (12 / 13) & (19 / 24)  \\
\hline
{ }  & Type I    &   205 / 465& 211 / 488& 216 / 442  & 237 / 503     \\ 
{ }&{ } & (159 / 170) & (138 / 151) & (67 / 70) & (81 / 87)  \\
Photometric & Type II     &     91 / 198   & 96 / 207  & 115 / 212 & 118 / 241   \\ 
{ }&{ } & (13 / 13) & (14 / 16) & (5 / 5) & (11 / 12)  \\
{ } &Galaxy     &     478 / 1048 & 564 / 1196  & 637 / 1176 & 663 / 1357    \\ 
{ }&{ } & (39 / 46) & (36 / 42) & (16 / 24) & (25 / 32)  \\
\hline 
X-ray&  XR I   &  368 / 793  & 395 / 837  & 450 / 814  & 455 / 907     \\ 
{ }&{ } & (153 / 161) & (133 / 149) & (63 / 65) & (81 / 87) \\
{ }&XR II   &    293 / 636  &334 / 731   & 351 / 702 & 396 / 837   \\ 
{ }&{ } & (50 / 58) & (49 / 53) & (20 / 25) & (32 / 37)  \\
\hline
Radio&RL&   106 / 232 & 116 / 243 & 140 / 269 & 136 / 287    \\ 
{ }&{ } & (18 / 18) & (16 / 18) & (8 / 9) & (10 / 11)  \\
{ } &RQ    &   185 / 408   &183 / 392  & 231 / 436 & 210 / 439  \\ 
{ }&{ } & (118 / 128) & (109 / 121) & (58 / 65) & (71 / 77)  \\
\hline
\hline
Total &{ } & 777 / 1715  & 874 / 1895 &    971 / 1835  & 1021 / 2107  \\
{ }&{ } & (213 / 231) & (190 / 211) & (88 / 99) & (117 / 131)  \\
\hline
\hline

\end{tabular}
\end{table*}

\subsection{Variability properties}\label{var_prop}

\begin{figure*}
\gridline{\fig{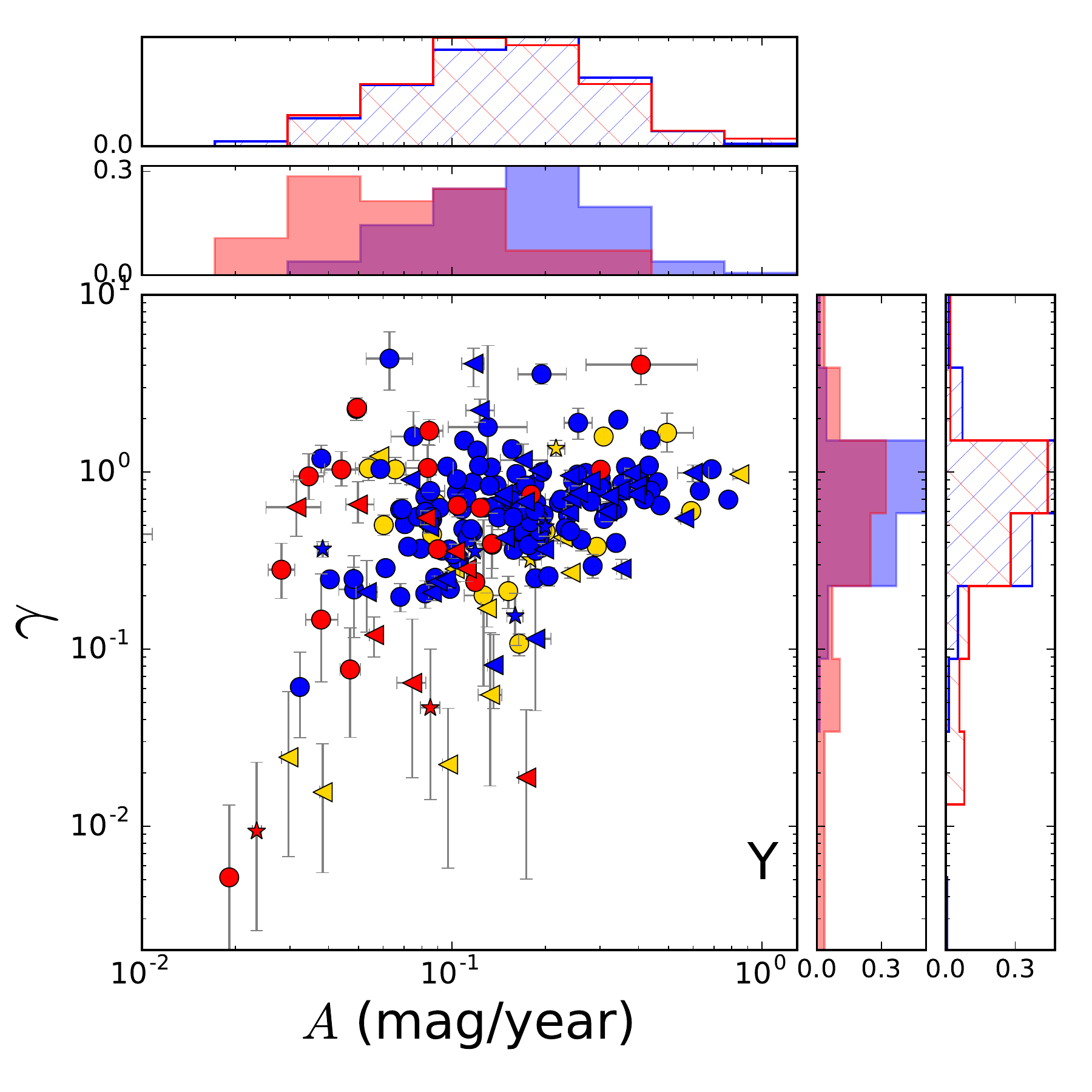}{0.5\textwidth}{}
          \fig{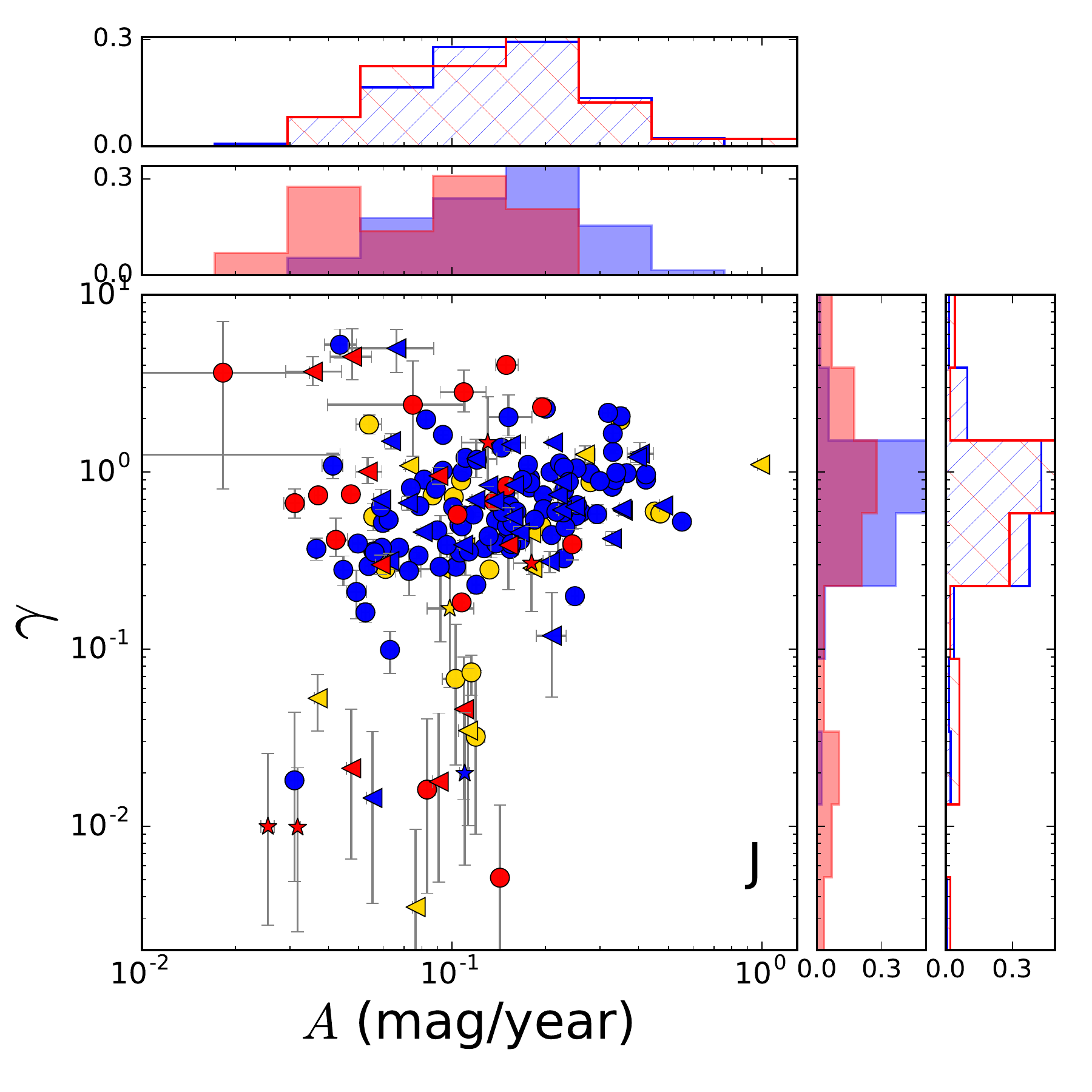}{0.5\textwidth}{}
          }
\caption{Distribution of the Structure Function power law parameters $A$ and $\gamma$, for the variable sources, in the Y and J bands, in logarithmic space. The blue sources correspond to BL AGN, the red sources to NL AGN, and the yellow to sources without spectroscopic classification. The circles correspond to XR I AGN, the triangles to XR II AGN, and the stars to sources without X-ray classification. The error of the measurements are shown with grey error bars. For most of the sources, the size of the error bars is smaller than the marker size. Along the axes we show the projected $A$  and $\gamma$ distributions for the BL AGN (blue shaded), NL AGN (red shaded), XR I AGN (blue hatched), and  XR II AGN (red hatched).\label{figure:SF_all}}
\end{figure*}

Figure \ref{figure:SF_all} shows the distribution of the power law parameters of the Structure Function $A$ and $\gamma$, for the variable sources in the Y and J bands in logarithmic scale. We mark the sources according to their spectroscopic and X-ray classifications. Histograms at the top and right hand side of the plots better represents the normalized distributions of variable sources of different AGN populations.

The mean and 1$\sigma$ errors of the Structure Function parameters for the variable sources are:  $A_{Y}=0.15_{-0.07}^{+0.16}$ and $\gamma_{Y}=0.62_{-0.32}^{+ 0.42}$ for the Y band, and  $A_{J}=0.13_{-0.09}^{+0.12}$ and $\gamma_{J}=0.63_{-0.29}^{+0.56}$. Clearly several sources have values of $\gamma$ consistent with DRW process. However, there are some sources with $\gamma >1.0$ (36 and 43 for the Y and J bands, respectively), which implies deviations from a DRW process ($\gamma=0.5$). These results are consistent with previous analysis, which have found that CARMA models with higher orders (and not a simple DRW model) better describe AGN light curves (e.g. \citealt{Kasliwal15,Kasliwal17,Simm16}).

From Figure \ref{figure:SF_all} we can see that the distribution of the $A$ parameter has a noticeable difference when we compare the sources classified as BL and NL, in particular in the Y band. On the other hand, when we compare the XR I and XR II sources the difference is less evident. In order to have a more quantitative comparison of the Structure Function parameters distributions, we performed a two-sample Anderson-Darling test \citep{Anderson-Darling} for the $A$ and $\gamma$ parameters considering the spectroscopic and X-ray classification. Since the Anderson-Darling test does not take into account the errors of the parameters, we only considered in the test those variable sources with a measured parameter having a signal to noise ratio higher than 3, considering as the error of the parameter the average of the lower and upper errors given by the Bayesian analysis. According to the test, for the BL and NL, the distributions of the $A$ parameter are different at a 99.5\% significance level in both Y and J bands, with a $p_{value}$ of $8.99\times10^{-6}$ and $1.69\times10^{-4}$ for the Y and J respectively. For the XR I and XR II sources, no statistically significant difference is found for the $A$ parameter, with a $p_{value}$ of $1$ and $0.87$ for the Y and J, respectively. For the case of the $\gamma$ parameter, no statistically significant differences are found in any filter for the spectroscopic and X-ray classifications. The $p_{value}$ are $0.052$ and $0.076$ for the Y and J bands in the spectroscopic classification, and $0.48$ and $0.16$ for the Y and J bands in the X-ray classification.

Since the dynamic range in redshift considered in this analysis is wide, we repeated the analysis selecting the smallest bin of redshift where we can ensure the presence of at least 6 sources belonging to every population (BL, NL, XRI and XRII), in order to have confident results from the Anderson-Darling test. This requirement is accomplished by the bin of redshifts between 0.3 and 1 (see Figure \ref{figure:hist_redshift}). In this bin we expect to observe emission coming from the accretion disk in both Y and J bands. The results of the Anderson-Darling test are the same. The distributions of the $A$ parameter for the spectroscopic classification are different at a 99.5\% significance level in both Y and J bands, with a $p_{value}$ of $0.01$ and $0.03$ for the Y and J, respectively. For the case of the $A$ parameter in the  X-ray classification ($p_{value}$ of $0.22$ and $0.1$) and the $\gamma$ parameter in both classifications (spectroscopic: $p_{value}$ of $0.41$ and $0.40$; X-ray: $p_{value}$ of $0.39$ and $0.23$), no statistically significant differences are found in any filter.

Figures \ref{figure:lc_fig} and \ref{figure:lc_fig2} show examples of light curves variable in the four photometric bands. We plotted the light curves of these two sources because they are particularly interesting. In Figure \ref{figure:lc_fig} the source is classified as BL - XR II. On the other hand, in Figure \ref{figure:lc_fig2}, the source is classified as NL - XR II. Even though this source is NL and obscured in the X-ray regimes, we can detect its variation in the four photometric bands. Its variability parameters in the Y filter band are: $A=0.05$ (mag/year) and $\gamma=0.69$. This source could potentially be an example of ``Changing look" AGN.

\subsection{Dependency with redshift and Luminosity}\label{z_dependency}

Redshift will obviously change the rest frame emission observed by each band. Any correlation with redshift, therefore, needs to take this into account. Besides, previous analysis have shown evidence of an anti-correlation between optical/UV luminosity with the variability amplitude (e.g. \citealt{Uomoto76,Hook94,Trevese94,Cristiani97,Wilhite08,Kelly09,MacLeod11,Meusinger13,Simm16,Caplar17} ). In other words, a more luminous (and probably larger) system varies, at a given fractional amplitude, on larger time scales.

In order to test any possible correlation of the Structure Function parameters with luminosity, we used bolometric luminosities ($L_{BOL}$) from \cite{Lusso12}. We also test any correlation with the intrinsic Hard X-ray luminosity ($HL_{int}$) (i.e., rest frame luminosity corrected by absorption), using the X-ray data from \cite{Marchesi16}. The value of $HL_{int}$ was computed using the rest frame observed luminosity ($HL_{obs}$) and the luminosity absorption correction ($k_{corr}$) provided by \cite{Marchesi16}:  $HL_{int}=HL_{obs}/k_{corr}$. We tested the correlations with luminosity and redshift for those variable sources with measured values of both $L_{BOL}$ and $HL_{int}$, which corresponds to 139 and 123 sources in the Y and J bands, respectively.

We calculated the Spearman rank-order correlation coefficient $r_s$ for log$(A)$ vs log$(1+z)$  and for log$(\gamma)$ vs log$(1+z)$ in the Y and J bands. We find clear evidence of positive correlation between $A$ and $(1+z)$. The values of the coefficient for the correlation between $A$ and $(1+z)$ are  $r_s=0.47$ ($p_{value}=5.5\times10^{-9}$) and $r_s=0.54$ ($p_{value}=1.4\times10^{-10}$) for Y and J respectively. For the case of  log$(\gamma)$ vs log$(1+z)$, the correlation is not evident. The value of the Spearman rank-order correlation coefficients are $r_s=0.08$ ($p_{value}=0.38$) and $r_s=0.1$ ($p_{value}=0.29$) for Y and J, respectively.

For the case of $L_{BOL}$, the Spearman coefficient showed some evidence of a positive correlation between log$(A)$ and log$(L_{BOL})$, contrary with what was expected from previous analysis. The values of the coefficient are $r_s=0.3$ ($p_{value}=3.7\times10^{-4}$) and $r_s=0.3$ ($p_{value}=6.6\times10^{-4}$) for Y and J, respectively. However, this result might be affected by the wide dynamic range in redshift considered in the sample(z $\sim$ 0.3 - 4). Thus, in order to disentangle whether the positive correlation is driven by redshift or by luminosity, further analysis is needed. Moreover, the correlation analysis showed no evidence of correlation between  log$(\gamma)$ and log$(L_{BOL})$. The coefficients are $r_s=0.08$ ($p_{value}=0.32$) and $r_s=-0.04$ ($p_{value}=0.65$) for Y and J, respectively.

Finally, for the case of $HL_{int}$, the correlation coefficient showed a weak (or non) positive correlation between log$(A)$ and log$(HL_{int})$, with $r_s=0.19$ ($p_{value}=0.024$) and $r_s=0.15$ ($p_{value}=0.093$) for Y and J, respectively. On the other hand, for  log$(\gamma)$ and log$(HL_{int})$, the correlation is negligible, with $r_s=0.01$ ($p_{value}=0.9$) and $r_s=0.01$ ($p_{value}=0.87$) for Y and J, respectively. Similar to the case of $L_{BOL}$, we need further analysis to say if the correlation between log$(A)$ and log$(HL_{int})$ is affected by the wide dynamic range in redshift.

To test whether the positive correlation of the amplitude of the Structure Function with luminosity can be due by a positive correlation with redshift, we calculated the correlation of $A$ with both $L_{BOL}$ and redshift in the logarithmic space, i.e., we computed $$\text{log}_{10}(A)=a \text{log}_{10}(L_{BOL}/10^{45} \text{erg s}^{-1}) + b  \text{log}_{10}(1+z) +c$$ for the same sources considered in the previous analysis. For this purpose, we computed the Weighted Least Squares linear regression (WLS), considering as weights the inverse of the variance of log$_{10}(A)$, calculated as $\sigma^2(\text{log}_{10}A)=(0.434*(A_{loerr}+A_{uperr})/2A))^2$. A summary of the regression for the Y and J filter bands can be found in Table \ref{tab:regression}. From the table, we can see that the correlation between log$(A)$ and log$(L_{BOL})$ is in fact negative, and it is statistically significant but weak, and the correlation between  log$(A)$ and log$(1+z)$ is positive and significant. Therefore, the positive correlation between log$(A)$ and log$(L_{BOL})$ obtained using the  Spearman coefficient was actually an effect of the positive correlation with redshift. Thus, whenever we perform a correlation analysis for the luminosity, we need to consider the redshift of the source as a second independent variable.

In Table \ref{tab:regression} we can also see the results for the regression of $A$ with both $HL_{int}$ and redshift in the logarithmic space. In this case, we computed $$\text{log}_{10}(A)=a \text{log}_{10}(HL_{int}/10^{43} \text{erg s}^{-1}) + b  \text{log}_{10}(1+z) +c$$
The results of the WLS analysis shown that the correlation with $HL_{int}$ is not statistically significant in any band. Therefore the weak positive correlation observed in the previous analysis might be produced by the positive correlation with redshift.

\begin{table}[h!]
\renewcommand{\thetable}{\arabic{table}}
\centering
\caption{Results of the WLS. $p_{values}$ in brackets.}\label{tab:regression}
\begin{tabular}{ccccc}
\tablewidth{0pt}
\hline
\hline
Filter & luminosity & a  & b  & c   \\
&  &  luminosity & redshift &  intercept \\
\hline
\hline
Y& $L_{BOL}$  &  $-0.11 \pm 0.05$ & $1.38 \pm 0.29$  & $-1.22 \pm 0.1$  \\ 
 & & (0.035) & (0.0)  &  (0.0) \\
Y& $HL_{int}$  &  $-0.05 \pm 0.03$ & $1.14 \pm 0.26$  & $-1.12 \pm 0.1$  \\
 & & (0.075) & (0.0)  &  (0.0) \\
J& $L_{BOL}$  &  $-0.13 \pm 0.05$ & $1.85 \pm 0.26$  & $-1.42 \pm 0.09$  \\
 & & (0.018) & (0.0)  &  (0.0) \\
J& $HL_{int}$  &  $0.003 \pm 0.03$ & $1.52 \pm 0.24$  & $-1.38\pm 0.09$  \\
 & & (0.9) & (0.0)  &  (0.0) \\
\hline
\hline
\end{tabular}
\end{table}

\begin{figure}
\begin{center}
\includegraphics[scale=0.4]{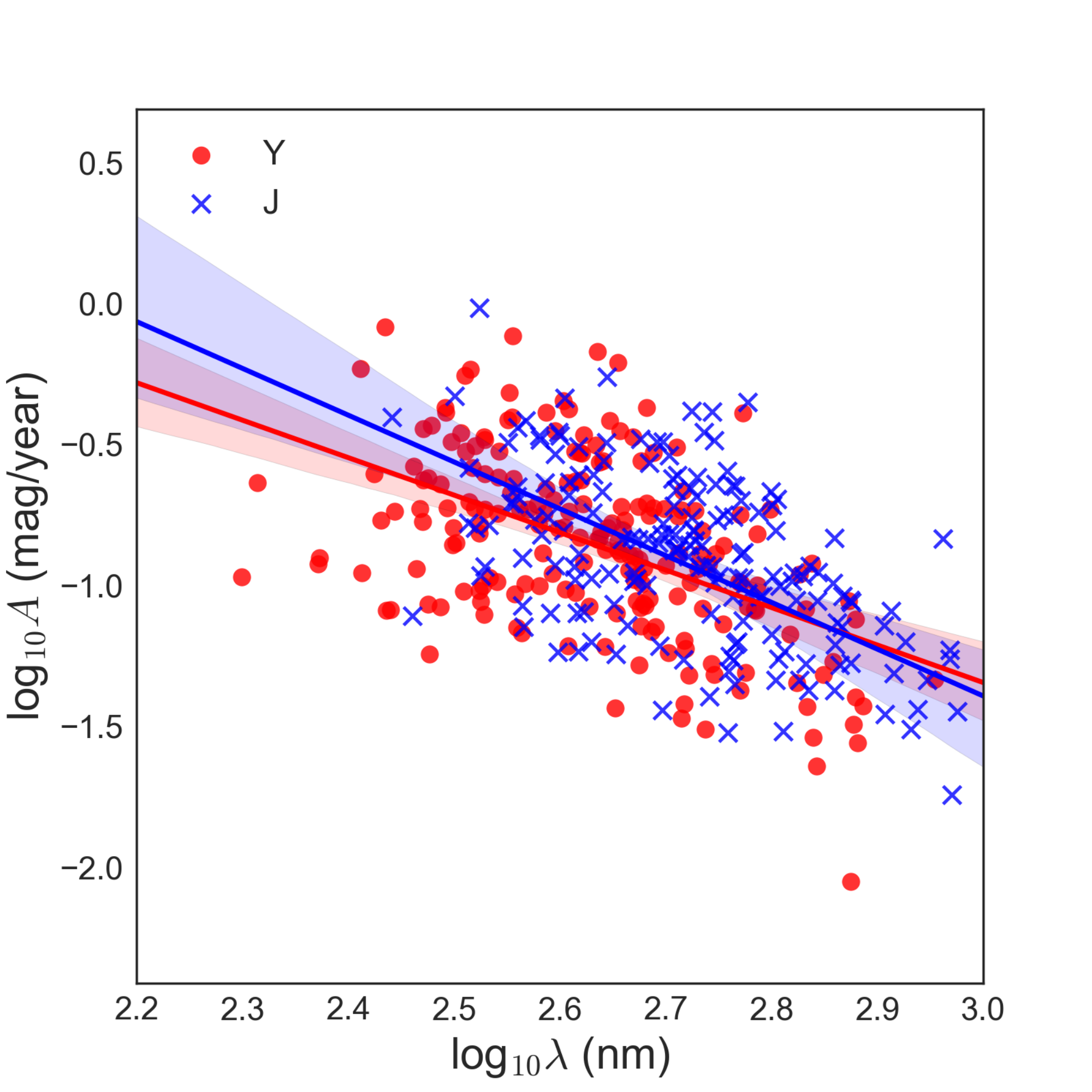}
\caption{Amplitude of the Structure Function $A$ vs rest frame wavelength of emission ($\lambda$) at logarithmic scale for the variable sources, in the Y (red circles) and J (blue crosses) bands. The blue  and red lines show the linear regression between log$_{10}(A)$ and log$_{10}(\lambda)$ for the Y and J bands respectively. The shaded regions show the $95 \%$ error of the regression.\label{figure:A_vs_z}}
\end{center}
\end{figure}

The positive correlations with redshift are consistent with observing bluer regions of the AGN SED as redshift increase, as bluer emission is expected to vary with larger amplitude. Figure \ref{figure:A_vs_z} shows $A$ vs rest frame wavelength of emission ($\lambda$) at logarithmic scale for the variable sources, in the Y and J bands. There is a clear anti-correlation between $A$ and $\lambda$, and the result of the linear regression for the Y and J bands are consistent at $95\%$ level. However, there is a large dispersion in the correlation. This can be related to other properties of AGN that may affect the amplitude of the variability aside from the emission wavelength, like the bolometric luminosity (which, as we already demonstrated, anti-correlates with the amplitude of the variability), the black hole mass, the accretion rate, among other physical properties. As well as the length and quality of the light curves, among observational factors.

The lack of correlation between $\gamma$ and redshift might indicate that the structure of the variability is independent of wavelength. This is consistent with previous analysis of optical and NIR light curves that claim that light curves of AGN observed at different wavelengths have the same structure or shape in time scales of months to years, but showing time lags between them, due to the distance between the emitting regions, and showing a decrement in the amplitude of the variation \citep{Lira11,Lira15}.

\begin{figure}
\begin{center}
\includegraphics[scale=0.35]{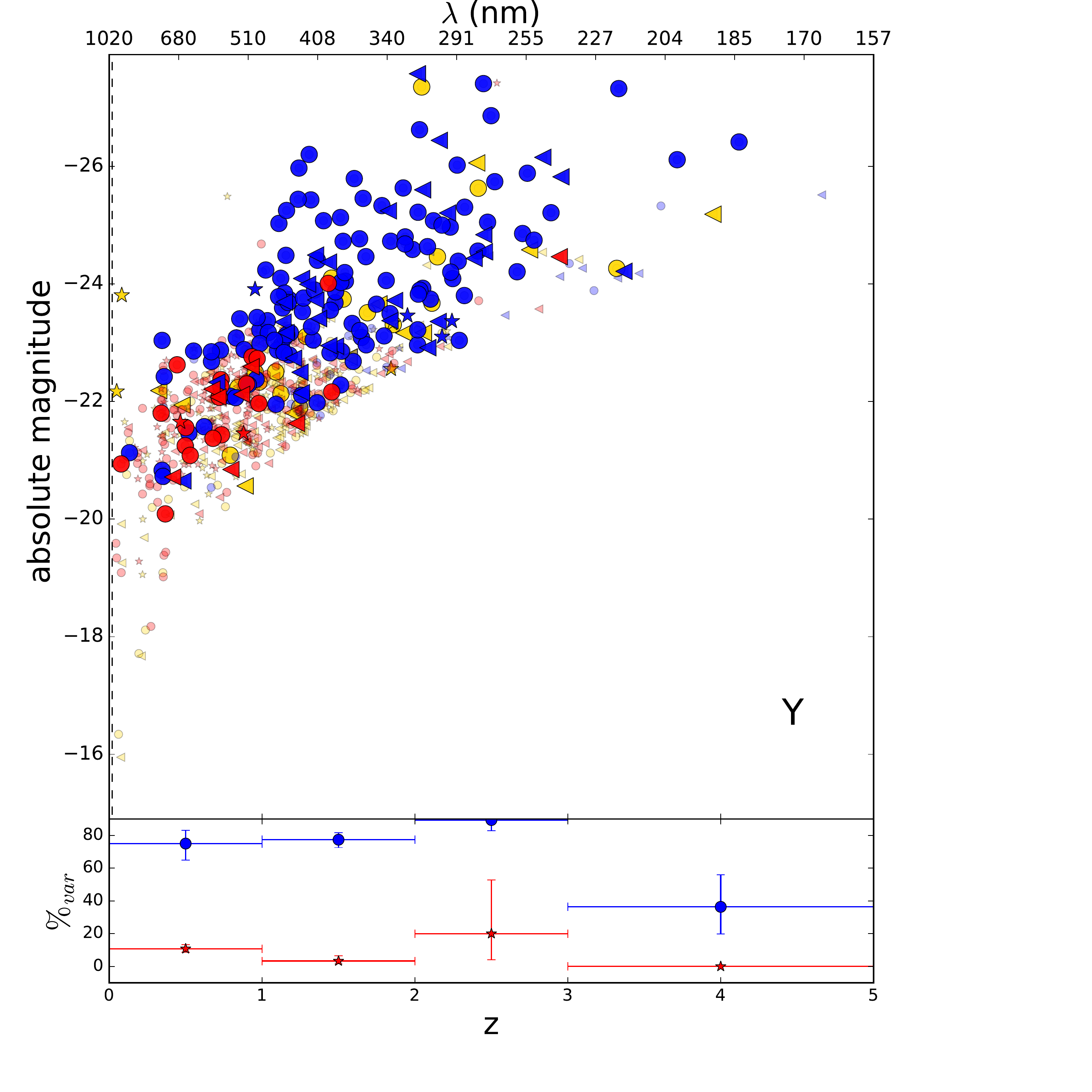}
\caption{Top: absolute magnitude (no k-corrected) vs redshift for the Y band. The small shaded sources are non-variable. The blue sources corresponde to BL AGN, the red sources to NL AGN, and the green to sources without spectroscopic classification. The circles corresponde to XR I AGN, the triangles to XR II AGN,and the stars to sources without X-ray classification. As a reference, we include in the top x axis the value corresponding to the rest frame wavelength of emission. Bottom: fraction of variable BL (blue) and NL (red) sources. \label{figure:z_vs_mag}}
\end{center}
\end{figure}

Figure \ref{figure:z_vs_mag} shows the absolute magnitude for the variable sources (no k-corrected) vs redshift (bottom axis) and vs rest frame wavelength of emission, for the Y band (top axis). In the figure we mark those sources according to their spectroscopic and X-ray classifications, as before. At the bottom of the figure we show the fraction of variable BL and NL sources in logarithmic scale. 
The number of variable NL sources detected at $0.5 \mu$m $<\lambda_{rest}<1 \mu$m is low ($\lesssim 10\%$). However, since this wavelength range is expected to be dominated by emission from the accretion disk and not directly observable in most of the obscured systems, we would not expect to detect high variability from NL sources in this regime. These sources, however do not show the same variability properties as their BL counterparts, and cluster at significantly smaller values of $A$ and $\sigma$ in Figure \ref{figure:SF_all}. 

\begin{figure}
\begin{center}
\includegraphics[scale=0.46]{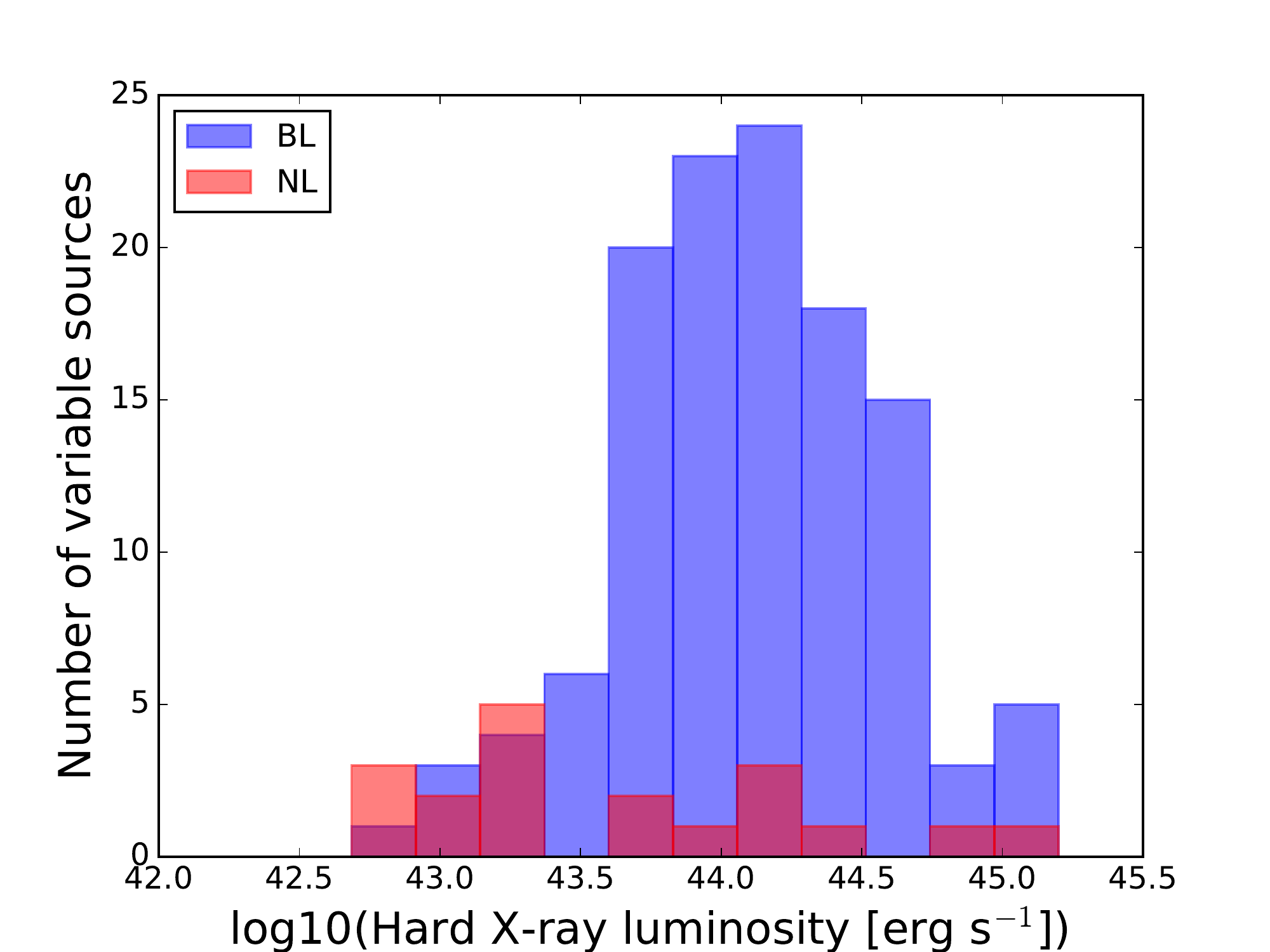}
\caption{Histogram of the intrinsic luminosity (rest frame and corrected for absorption) in the hard X-ray band, of variable sources in the Y band, considering the spectroscopic classification. BL are showed in blue and NL in red.\label{figure:hist_lum}}
\end{center}
\end{figure}

As torus emission is not expected around rest frame $0.5 \mu$m, the most likely explanation is that these variable NL sources correspond to BL AGN but where the host galaxy might be damping the variability signal (hence, yielding smaller values of the parameter $A$), and masking the presence of weak broad lines in the spectra.  One way to check this is by looking at the intrinsic X-ray luminosity in the hard band (rest frame luminosity and corrected for absorption), for variable BL and NL AGN. This is presented in Figure \ref{figure:hist_lum}. It clearly shows that most of the NL variable sources have low luminosity ($HL_{int}<10^{44}$ erg s$^{-1}$). We therefore discard ``True type II'' AGN as a possibility to explain our low luminosity variable NL sources, since we might expect to observe that their variability properties, like the Structure Function parameters, are similar to the properties of normal BL sources, which is not seen for most of our variable NL sources. For the case of the few bright NL variable sources, we cannot discard that they are ``True type II'' AGN.

\begin{figure}
\begin{center}
\includegraphics[scale=0.46]{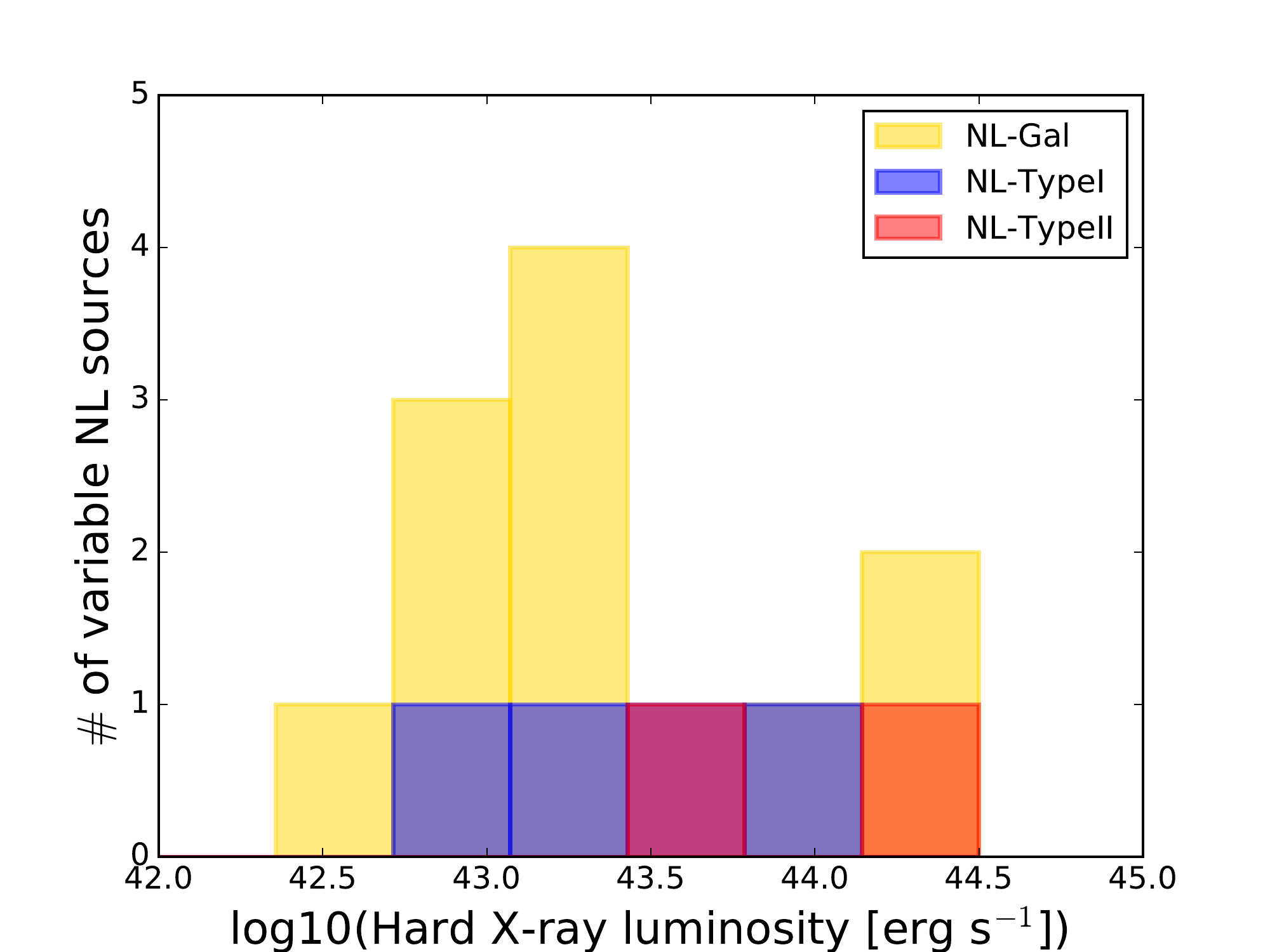}
\caption{Histogram of the intrinsic luminosity (rest frame and corrected for absorption) in the hard X-ray band, of variable NL sources in the Y band, split by their photometric classification. NL-TypeI are showed in blue, NL-TypeII in red, and NL-Gal in yellow.\label{figure:hist_NL}}
\end{center}
\end{figure}

 To have a better understanding of the nature of these variable NL sources, we plot in Figure \ref{figure:hist_NL} a histogram of $HL_{int}$ for the NL variable sources, split by their photometric classification. From the figure we can see that most of the low luminosity sources were adjusted by a galaxy template in \cite{Marchesi16}. This result supports our idea that most of the low luminosity variable NL sources correspond to BL sources whose emission is overshadow by their host galaxy. However, a few sources have a photometric classification of Type I. These sources can be either ``True type II'' AGN or normal BL whose optical spectrum does not sample the region where the broad lines are present, since the SED analysis reveals a continuum emission that is consistent with continuum emission of BL sources.

\subsection{Fraction of variable sources}\label{frac_var_sources}

\begin{figure}
\gridline{\fig{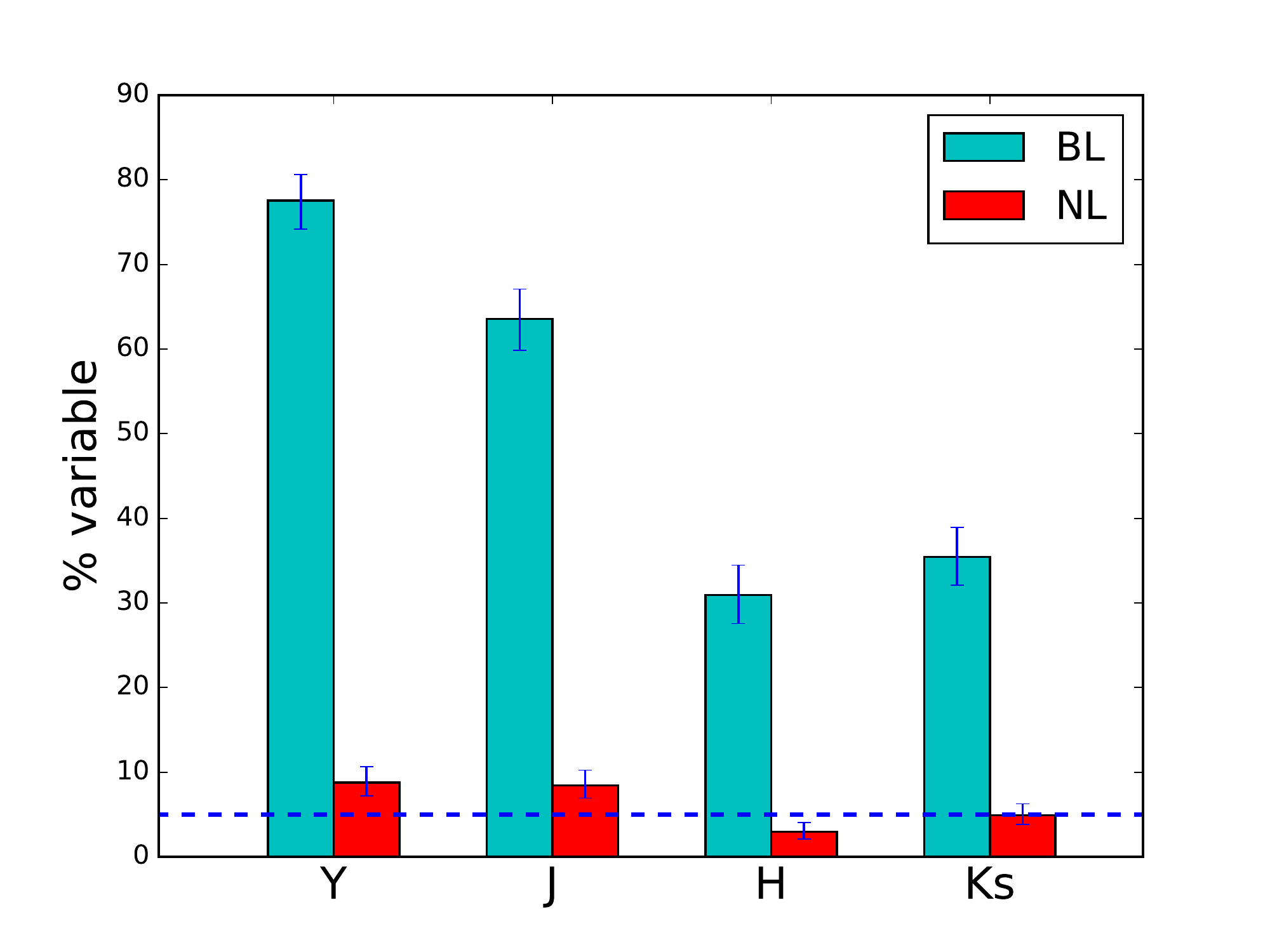}{0.48\textwidth}{(a)}}
\gridline{\fig{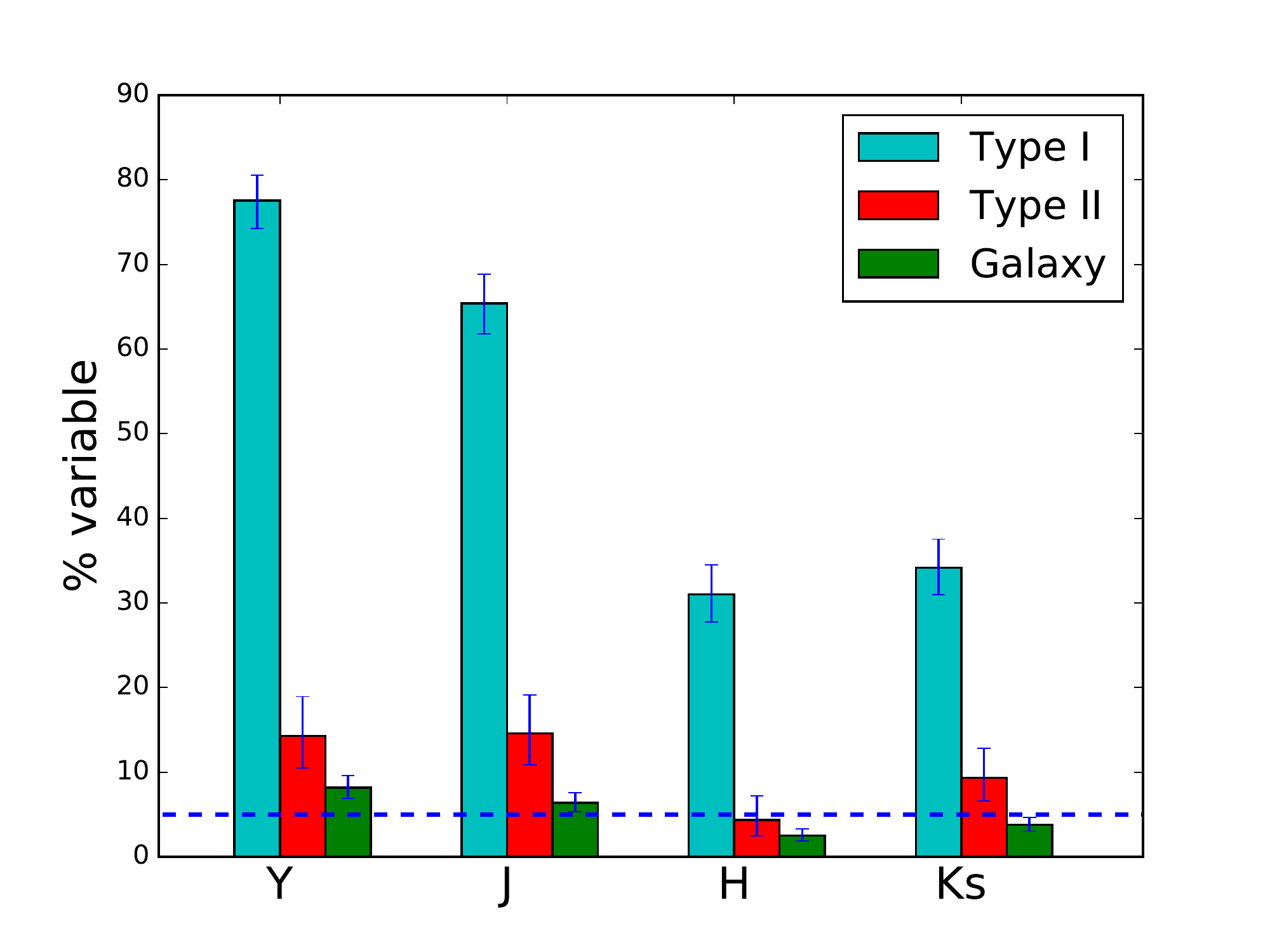}{0.48\textwidth}{(b)}}
\gridline{\fig{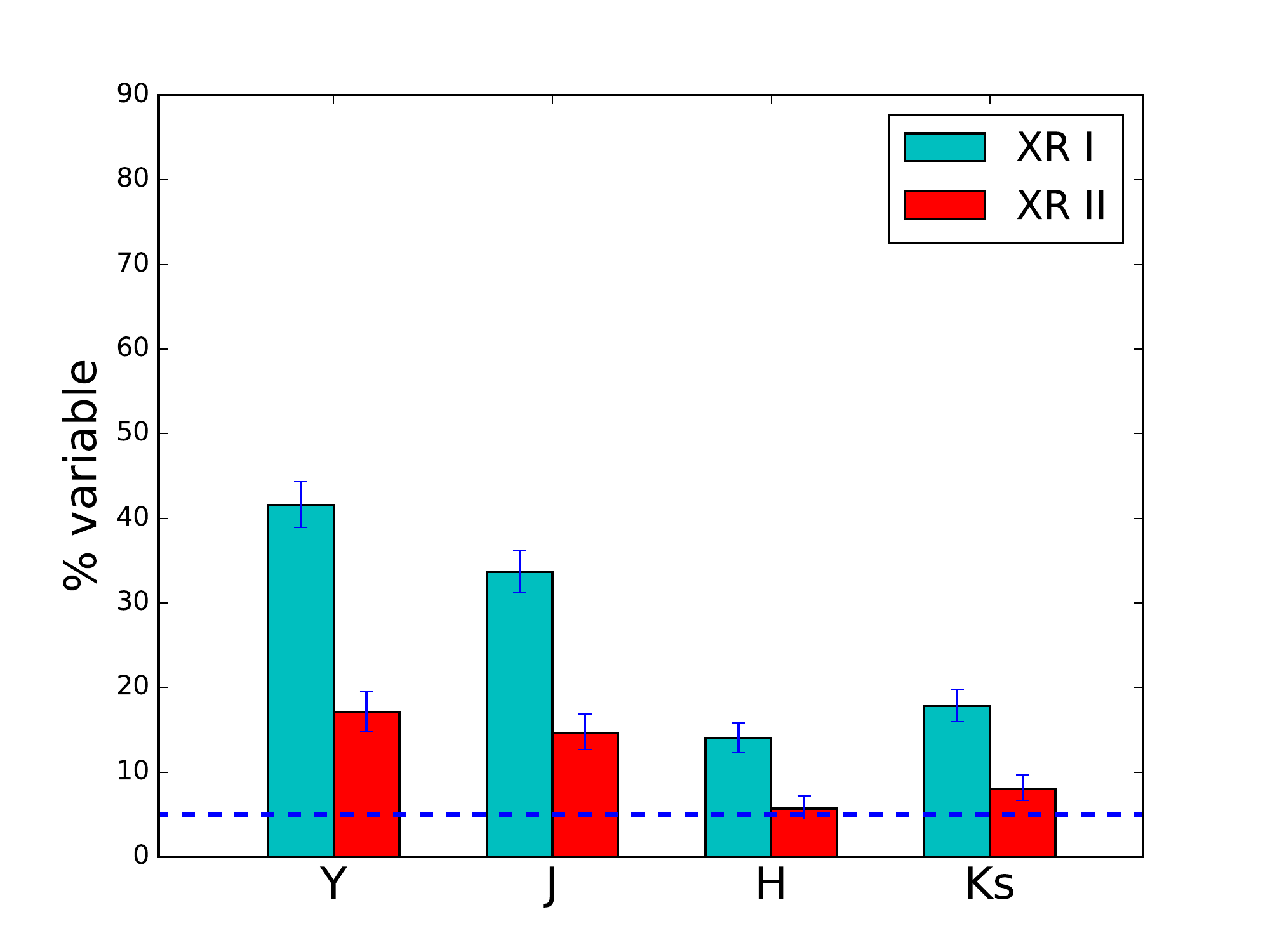}{0.48\textwidth}{(c)}}
\caption{Fraction of variable sources for the spectroscopic, photometric and X-ray classifications, for our four photometric bands. The blue dashed line demarks the 5\%, below this fraction we might expect to have several false positive variable sources. The error bars were calculated considering poisson statistic \citep{Gehrels86} \label{figure:fracvar}}
\end{figure}

In this section we show the results of our variability analysis considering the fraction of variable sources. The results shown in this section are for the four photometric bands. Figure \ref{figure:fracvar} (a) shows the fraction of variable sources for BL AGN and NL AGN (also see Table \ref{tab:summary_lc}). We can see that BL AGN have a higher fraction of sources with detected variability in comparison with NL AGN. Besides, we can see a decrement in the fraction of variable BL AGN as we move to redder bands. This can be related to the increment of the photometric errors from the Y to Ks bands. However, we also have to consider that for centrally driven variations it is expected that the amplitude of the variability is lower at longer wavelengths, and therefore we might expect to have a reduction in the fraction of variable sources detected as we move to longer wavelengths. This is in fact observed in BL AGN, where the fraction of variable sources reduces systematically from the Y to the H band. The increase observed in the Ks band could be accounted by the presence of the torus, which has a  large solid angle as seen by the innermost region of the disk, hence boosting its variability.

 For the case of NL AGN we observe that the fraction of variable sources is above the $5\%$ only for the Y and J bands. In the context of the unified model, we might expect a very low probability to detect variable sources for the Y and J bands, and an increment in the fraction for the H and Ks bands. However, some optical variability might be expected for NL sources considering that the obscuring material is a distribution of moving clumps or clouds. Besides, as mentioned in section \ref{spec_class}, the classification of NL sources considers the lack of visible broad emission lines.  For the case of BL sources located at redshift $\sim 1$, depending of the spectral coverage, the typical broad line components might be out of the spectra, and therefore be classified as NL. For these sources, we might expect to detect optical variability.

For the case of the photometric classification, we show in Figure \ref{figure:fracvar} (b) the fraction of variable sources in the four photometric bands (also see Table \ref{tab:summary_lc}). We can see that the highest fraction of variable sources is for the Type I objects. They have a similar fraction of variable sources than BL AGN. The fraction of Type II variable sources is lower than Type I sources, but slightly larger than that of NL AGN. Sources classified as Galaxies show the lowest fraction of variability. It is important to notice that for most of sources best-fitted with a galaxy template by \cite{Marchesi16}, the X-ray luminosity is $>10^{42}$ erg/s, and therefore the sources are most likely AGN, although their optical-IR SED does not clearly show this. As before, variability is revealing unobscured AGN in sources where a SED analysis predicts otherwise.

We show in Figure \ref{figure:fracvar} (c) the fraction of variable sources for the X-ray classification (also see Table \ref{tab:summary_lc}). From the figure we can notice that XR I objects have a larger fraction of variable sources than XR II objects. Another obvious result is the lower fraction of X-ray classified variable sources when compared with those with an optical classification. This is not a luminosity effect. In fact, the mean of the X-ray luminosity of the X-ray sample is $10^{43.4 \pm 0.7}$ erg s$^{-1}$, while the same value for the sub-sample with optical spectroscopy corresponds to $10^{43.5 \pm 0.7}$ erg s$^{-1}$. We also checked whether this result is related to the method used to separate obscured and unobscured X-ray sources. We repeated the analysis separating the sources by their Hardness ratio. We considered a source as unobscured in X-rays if $HR<-0.2$. We obtained similar results than the ones showed in Figure \ref{figure:fracvar} (c). Therefore this result is not produced by the definition of X-ray obscuration used.

\begin{figure}
\gridline{\fig{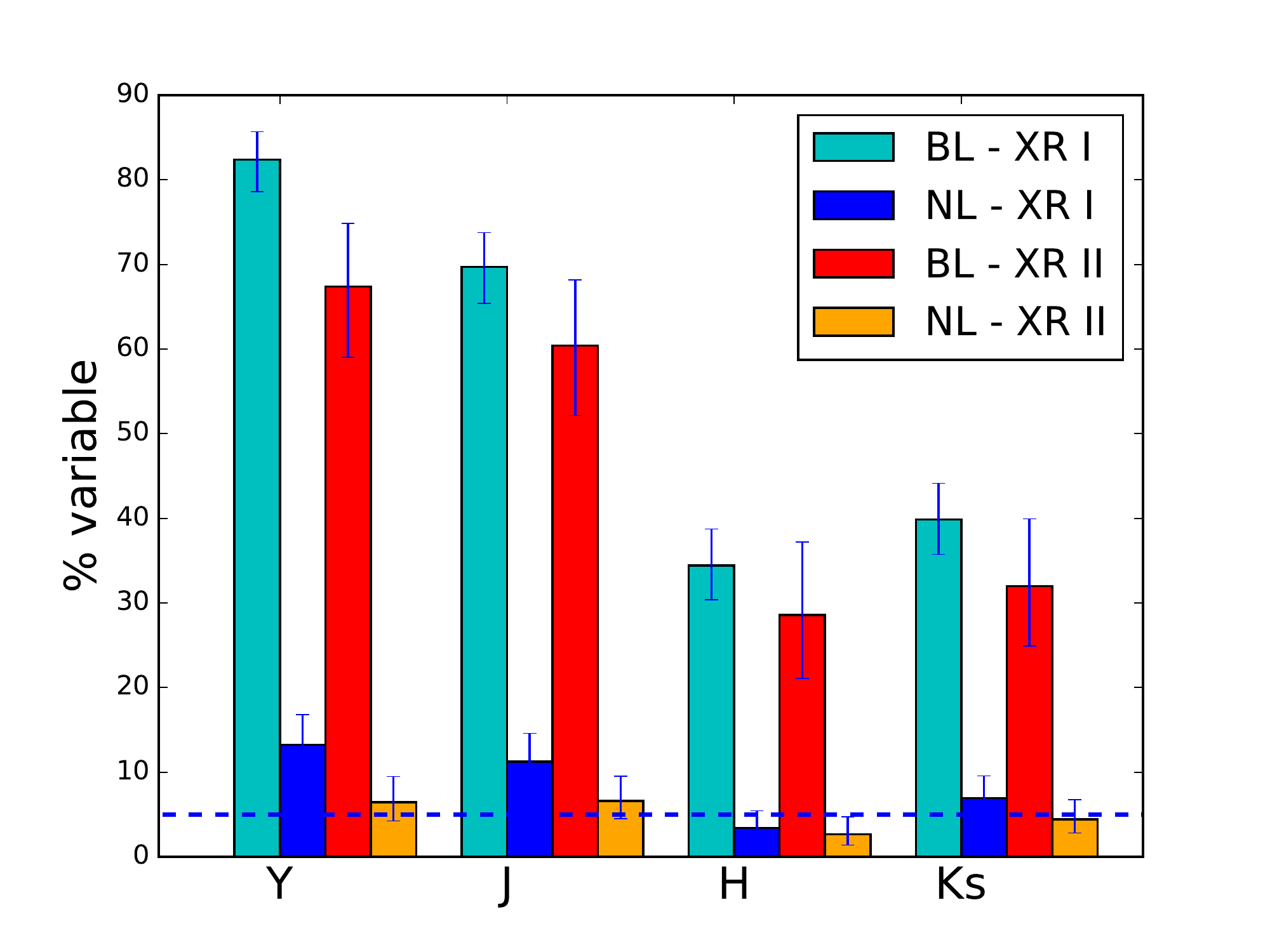}{0.48\textwidth}{(a)}}
\gridline{\fig{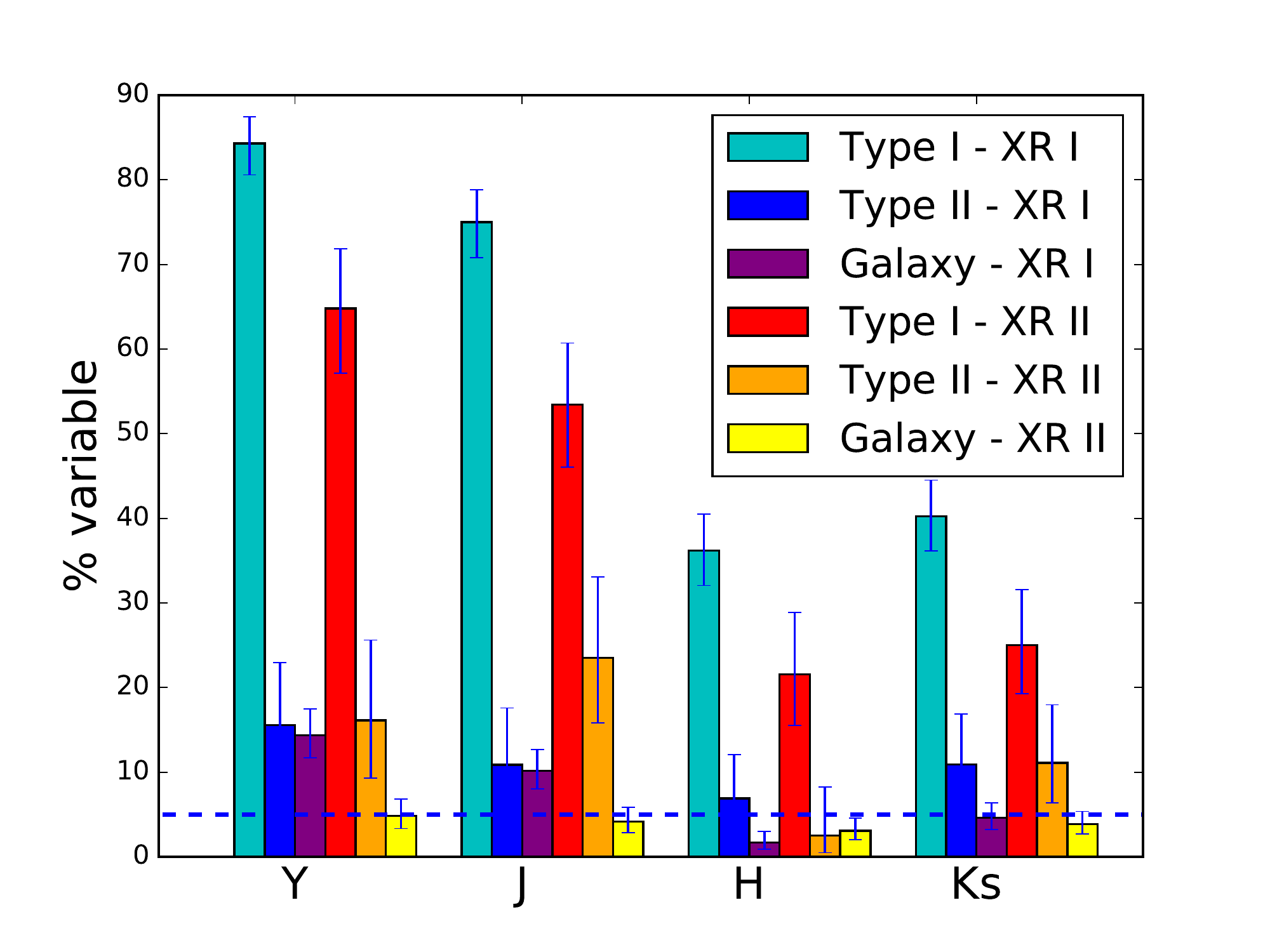}{0.48\textwidth}{(b)}}
\caption{Fraction of variable sources for the spectroscopic/X-ray classification, and for the photomeric/X-ray classification. The blue dashed line demarks the 5\%, below this fraction we might expect to have several false positive variable sources. The error bars were calculated considering poisson statistic \citep{Gehrels86} \label{figure:fracvar_2class}}
\end{figure}

To understand in a better way this difference, we plot in Figures \ref{figure:fracvar_2class} (a) and (b) the fraction of variable sources, considering the X-ray and spectroscopic classification, and the X-ray and photometric classification, respectively. From the figures, we can see that the difference in the fraction of variable sources is more closely related with the optical obscuration of the sources than with the X-ray obscuration. Sources unobscured in the optical range are the ones with the largest fraction of variable objects, irrespective of their X-ray classification. In fact, sources classified as unobscured in the X-rays but obscured in the optical range (NL - XR I or Type II - XR I) have a lower fraction of variable objects than sources classified as obscured in the X-rays and unobscured in the optical range (BL - XR II or Type I - XR II). As a significant fraction of XR I sources have a NL classification (136/368,142/395, 175/450 and 159/455, in the Y to K bands, respectively), this explains the low fraction of variable sources classified as unobscured in X-rays. We repeated the analysis separating the sources by their HR. Again, we obtain similar results, and therefore our result are not biased by the definition of X-ray obscuration.  We also repeated the analysis for the redshift bin $z\sim 0.3-1.0$, and the results were consistent with what we found for the whole sample.

These results might be related to the differences in the origin of the obscuration in the optical and X-ray regimes. \cite{Marchesi16b} analyzed the X-ray spectral properties for a sub-sample of the Chandra COSMOS-Legacy Survey catalog, and showed that most of the sources classified as BL - XR II have $L_{2-10\text{keV}}>10^{44} \text{erg s}^{-1}$. They conclude that the existence of these objects suggests that optical and X-ray obscuration can be caused by different mechanisms, and that the X-ray obscuration might be due to dust-free material surrounding the inner part of the nuclei. \cite{Merloni14} presented a detailed discussion on the nature of sources that have inconsistent classification in the optical and X-ray regimes. They found that sources classified as BL - XRII (type-12 in \citealt{Merloni14}) tend to have higher luminosities. Additionally, they showed that the main differences between sources classified as BL - XRII and sources classified as BL-XRI (or type-11) happen in the X-ray regime. They demonstrated  that the optical spectra and the SED of these two populations do not show substantial differences, and proposed that the excess absorption in the X-ray regime presented in the type-12 sources could be produced by dust-free material within (or inside) the broad line region. Our variability analysis strengthens this idea, since we do not observe significant differences in the variability features and in the fraction of variable sources between the BL - XRI and BL - XRII populations.

\cite{Marchesi16b} showed that most of the sources classified as  NL - XR I have $L_{2-10\text{keV}}<10^{43} \text{erg s}^{-1}$, and they expected a fraction of these sources to be ``True type II'' AGN. \cite{Merloni14} found that the sources classified as  NL - XR I (or type-21) tend to have lower luminosities, $L_{X}<10^{44} \text{erg s}^{-1}$, and that their composite spectrum and SED reveals evidence of host galaxy dilution. They proposed that a minority of these objects could be ``True type II'' AGN. Our variability analysis agrees with the findings of  \cite{Merloni14}, since of the 18 sources classified as variable NL - XR I in the Y band, 10 have photometric classification of galaxy, 4 are Type II, and 4 are Type I. Therefore, these 10  NL - XR I - Galaxy sources can be a clear example of BL sources with host galaxy dilution, but whose variability can be detected. The 4 NL - XR I-Type I sources can be examples of ``True type II'' AGN or BL AGN whose optical spectrum does not cover the region where the broad line are present, since their Structure Function features are in agreement with those of BL sources ($A=[0.12,0.13,0.18,0.12]$, $\gamma=[0.63,0.37,0.76,0.67]$), and  $z=[1.43,0.98,0.73,0.74]$. Finally, the 4 sources classified as variable NL-Type II, probably are examples of optically obscured variable sources.

\begin{figure}
\begin{center}
\includegraphics[scale=0.45]{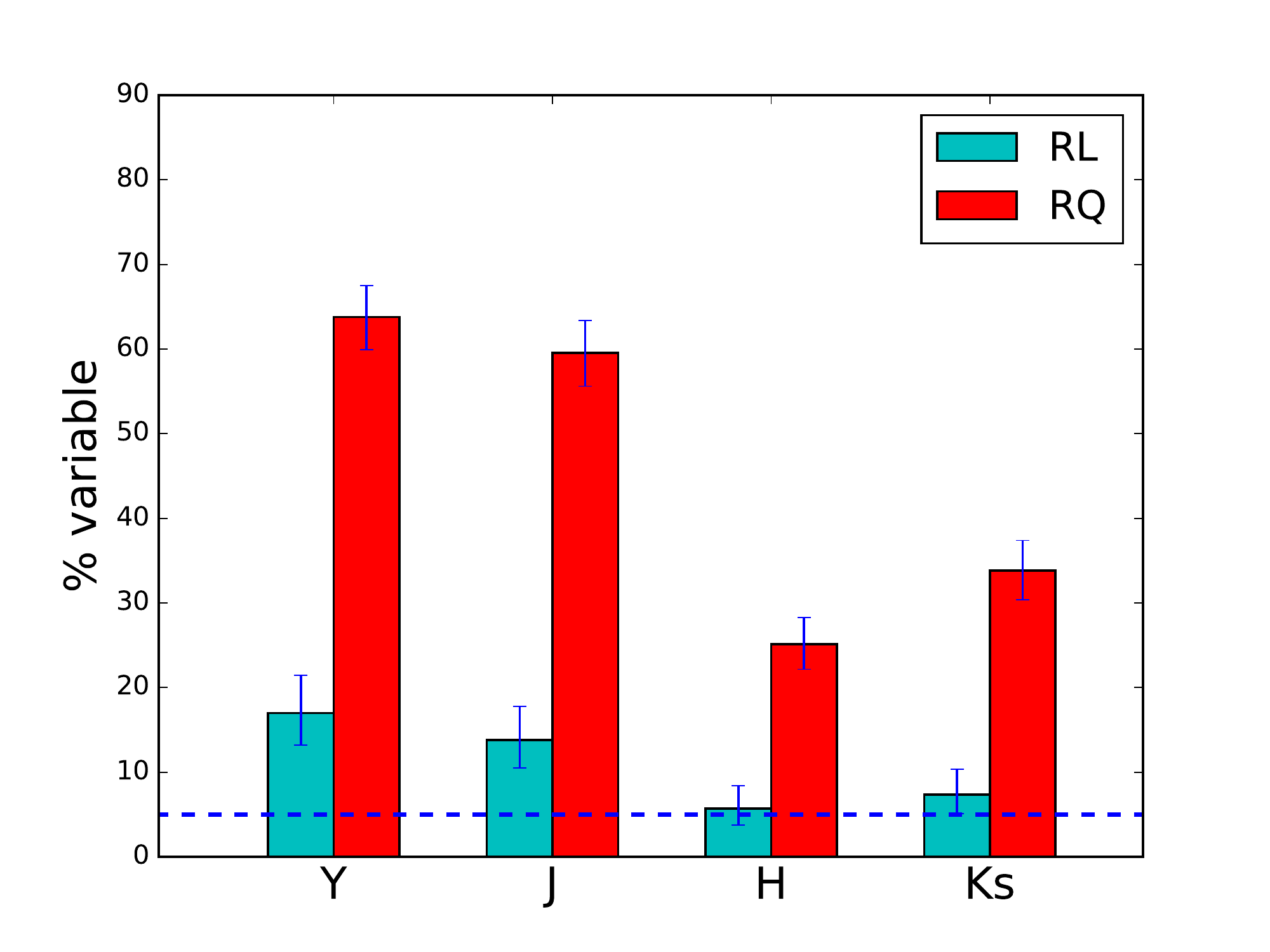}
\caption{Fraction of variable sources for the radio classification, for our four photometric bands. Radio Loud (RL) objects are showed in cyan, and Radio Quiet (RQ) in red. The blue dashed line demarks the 5\%, below this fraction we might expect to have several false positive variable sources. The error bars were calculated considering poisson statistic \citep{Gehrels86}.\label{figure:fracvar_radio}}
\end{center}
\end{figure}

Figure \ref{figure:fracvar_radio} shows the fraction of variable sources for the Radio Classification. In this case, the Radio Quiet sources show a considerably higher fraction of variable sources than the Radio Loud sources. A $72\%$ of the RL and  a $60\%$ of the RQ sources are classified as NL AGN. The large fraction of RL objects with a NL classification is not surprising. RL objects represent a distinctive AGN population not only because of their radio properties. They are also characterized by massive hosts, large BH masses and very low accretion rates (\citealt{Heckman14}, and references therein). These traits seem to suggest that these are systems at the end of their life cycles of actively growing BH masses. The very low Eddington ratios in turn make accretion highly inefficient and the optically thick, geometrically thin disks usually invoked in most AGN would be replaced by an advection-dominated or radiatively inefficient accretion flow (ADAFs/RIAFs, \citealt{Narayan94,Narayan95,Blandford99}). The absence of a classical accretion disk, and most likely of a classical BLR, would explain the spectroscopic classification of the RL sources and of the observed lack of variability. In this scenario, the lack of significant variability would not be due to obscuration but to the intrinsic nature of these extremely low accreting sources.

\section{Discussion and Conclusions}\label{discussion}

In this paper we present a characterization of AGN variability in the near infrared regime, using data from the UltraVISTA survey \citep{McCracken12}. UltraVISTA repeatedly imaged the COSMOS field in 5 bands (YJHKs and NB118), covering an area of 1.5 deg$^2$, to achieve a very deep final image. The survey provides excellent quality, high spatial resolution data (with a mean seeing of $\sim 0.8$") at different epochs, and has allowed us to analyze near-IR variability within a time span of almost five years with good sampling. Besides, the depth of the images allows us to cover a wide redshift range, accessing the optical and near-infrared rest-frame emission. We used four public catalogs \citep{Lusso12,Muzzin13,Marchesi16,Laigle16}  to construct a catalog of X-ray selected AGN, with information about their bolometric luminosities and spectroscopic, X-ray and radio properties. We used these catalogs to analyze the differences in the variability properties of sources classified according to their obscuration  in the X-ray and optical range, and according to their radio properties.

Our variability analysis is based mostly on sources classified as variable. When we downsized the sample by selecting only those sources with well sampled light curves (with more than 20 epochs and a length larger than 200 days in the rest frame), we only missed a small fraction of variable sources ($\sim 10\%$). While providing a much more significant fraction of variable sources. As we showed in Table \ref{tab:summary_lc}, the removed light curves are mostly non-variable.

According to the unified model, we might expect to detect a low fraction of NL variable sources in the optical range, since the obscuring material is expected to be a non-homogeneous distribution of moving clumps. It also predicts that as we move to the infrared regime, we should observe re-processed emission coming from the dusty torus, and therefore we should be able to detect damped variability for both BL and NL AGN. The photometric bands used in our analysis (YJHKs) allowed us to access optical and near infrared rest-frame emission, depending on the redshift of the source and the band considered. These two predictions are verified by our analysis in section \ref{frac_var_sources} and Figure \ref{figure:fracvar}. For the case of the Ks band, for sources located at redshifts lower than 1.15, we observed emission coming from the NIR in the rest frame, and therefore, we should expect to detect variability for obscured and unobscured sources, therefore increasing the fraction of variable sources. This is seen in our data.

Previous variability analysis have mostly been focused on unobscured sources, however there are some cases where optical variability has been reported for type II AGN. \cite{Choi14} used SDSS data to select AGN candidates by variability. They found that contrary to the AGN unification model prediction, two of their six type II candidates showed a non-negligible amount of optical variability. \cite{Cartier15} used data from the QUEST-La Silla AGN variability survey to study optical variability of BL and NL AGN. They show that 80\% of the BL and 21\% of the NL sources are classified as variable, and from their Structure Function analysis, they found that BL and NL AGN have different distributions on the SF parameter space. \cite{Simm16} used a sample of variable X-ray selected AGN from the catalog of \cite{Brusa10}, to study optical variability. They reported that 96\% of the sources were classified as type I, and they mention that 7 type II AGN in their sample were variable, which were not included in the analysis. \cite{Simm16} also found that the amplitude of the variability anti-correlates with the bolometric luminosity.

In section \ref{var_prop} we showed that the variable NL sources have different distributions of the Structure Function $A$ parameter compared to the variable BL sources. This result is in agreement with the results of \cite{Cartier15}. For the case of the XR I and XR II sources, the differences in the same parameter are not statistically significant. We think that the most likely explanation for the existence of variable NL sources in the optical rest-frame range is that most of them, which are also characterized by low AGN luminosities, correspond to BL AGN whose host galaxy is damping the variability signal, since most of them have a photometric classifications of Galaxy. We also proposed in section \ref{frac_var_sources} that the four sources with variability properties similar to BL, with photometric classification of Type I, unobscured in X-rays and with high luminosities, correspond to ``True type II'' AGN or to BL AGN whose spectrum does not cover the region where the broad lines are present. For the case of the 4 variable  sources classified as NL-TypeII sources in the Y band, we propose that they are examples of variable optically obscured sources.  

The regression analysis of the Structure Function parameters with redshift and $L_{BOL}$ (section \ref{z_dependency}) showed that $A$ -- related with the amplitude of the variability in a time scale of 1 year -- has a positive correlation with redshift, and a weak anti-correlation with  $L_{BOL}$. These results are in agreement with previous analysis (e.g. \citealt{Simm16,Caplar17}). On the other hand, the correlation between $A$ and $HL_{int}$ is not statistically significant. The correlation of the amplitude of the variability with redshift reveals an anti-correlation of the amplitude with the wavelength of emission. As we move to redder bands, the amplitude of the variability decrease (see Figure \ref{figure:A_vs_z}).

For the case of the parameter related with the structure of the variability, $\gamma$, the results of our analysis in section \ref{z_dependency} did not show any correlation with redshift, $L_{BOL}$ or $HL_{int}$. We also showed in section \ref{var_prop} that several sources have values of $\gamma$ consistent with DRW processes, however there are a non-negligible number of variable sources with $\gamma>1.0$ (36 for the Y band and 43 for the J band), which reveals  deviations from a simple DRW process.

We also showed in section \ref{frac_var_sources} that the fraction of variable sources unobscured in the X-ray is lower than the fraction of variable sources unobscured in the optical (in the spectroscopic and photometric classifications). We demonstrated that when we split the sources by their spectroscopic and X-ray classifications and by their photometric and X-ray classification (Figure \ref{figure:fracvar_2class}), the differences in the fraction of variable sources are given by optical obscuration (i.e. spectroscopic and photometric classifications) and not by X-ray obscuration (X-ray classification). In other words, optical rest frame variability is indifferent to  X-ray obscuration. We think that an explanation is that optical and X-ray obscuration are caused by different mechanisms, and that X-ray obscuration might be due to dust-free material surrounding the inner part of the nuclei, as it was proposed by \cite{Merloni14} and \cite{Marchesi16b}.

\vspace{0.5cm}

We thank Elisabeta Lusso for supplying her bolometric luminosity catalog. We also thank the referee, for a careful reading of the manuscript and comments that led to its improvement.
This work was based on data products from observations made with ESO Telescopes at
the La Silla Paranal Observatory under ESO programme ID 179.A-2005 and
on data products produced by TERAPIX and the Cambridge Astronomy Survey
Unit on behalf of the UltraVISTA consortium.
P.S. acknowledges support by CONICYT through “Beca Doctorado Nacional, A\~no 2013” grant \#21130441. PL acknowledges Fondecyt Grant \#1161184. RC acknowledge support from STFC grant ST/L000679/1 and EU/FP7-ERC grant No. [615929]. The research leading to these results has received funding from the European Research Council under the European Union's Seventh Framework Program (FP7/2007-2013)/ERC Grant agreement no. EGGS-278202.

\bibliography{bibliography}



\end{document}